\documentclass[useAMS,usenatbib]{mnras}
\usepackage{amsmath}
\usepackage{mathtools}
\usepackage{amssymb}
\usepackage{graphicx}
\usepackage{multirow}
\usepackage{color}
\usepackage{array}
\usepackage{graphicx}
\usepackage[para,online,flushleft]{threeparttable}

\title[Granger causality and AGN time lags]{Revealing the intrinsic X-ray reverberation lags in IRAS~13224--3809 through the Granger causality test}

\author[P. Chainakun et al.]{P. Chainakun$^{1,2}$\thanks{E-mail: \href{mailto:pchainakun@g.sut.ac.th}{pchainakun@g.sut.ac.th}}, N. Nakhonthong$^{1}$, W. Luangtip$^{3,4}$, A. J. Young$^{5}$\\
$^1$School of Physics, Institute of Science, Suranaree University of Technology, Nakhon Ratchasima 30000, Thailand\\
$^2$Centre of Excellence in High Energy Physics and Astrophysics, Suranaree University of Technology, Nakhon Ratchasima 30000, Thailand\\
$^3$Department of Physics, Faculty of Science, Srinakharinwirot University, Bangkok 10110, Thailand\\
$^4$National Astronomical Research Institute of Thailand, Chiang Mai 50200, Thailand\\
$^5$H.H. Wills Physics Laboratory, Tyndall Avenue, Bristol BS8 1TL, UK}
\date{Accepted XXX. Received YYY; in original form ZZZ}

\pubyear{2021}

\begin{document}
\label{firstpage}
\pagerange{\pageref{firstpage}--\pageref{lastpage}}
\maketitle

\begin{abstract}

The Granger causality is an econometric test for determining whether one time series is useful for forecasting another one with a certain Granger lag. Here, the light curves in the 0.3--1 keV (reflection dominated, soft) and 1.2--5 keV (continuum dominated, hard) bands of Active Galactic Nuclei (AGNs) are produced, and the Granger lags are estimated and compared to the traditional lag-frequency spectra. We find that the light curves in the hard band Granger-cause (lead) those in the soft band, whereas the obtained Granger lags could be interpreted as the intrinsic reverberation lags. Then, we extract the Granger-lag profiles from 14 {\it XMM-Newton} observations of IRAS~13224--3809, and find that the lags are significant in 12 observations. The majority of the obtained Granger (intrinsic) lags are $\sim 200$--500~s. With the IRAS~13224--3809 mass of $2 \times 10^{6}~M_{\odot}$, these lags correspond to the true light-travel distance of $\sim20$--50~$r_{\rm g}$. Hence, by assuming a lamp-post geometry and a face-on disc, this places the corona at $\sim10$--25~$r_{\rm g}$ above the central black hole. Moreover, multiple Granger lags consisting of the small and large lags of $<500$~s and $> 1000$~s are detected in 4 observations. This means that the corona height can significantly evolve from $\sim10$--25~$r_{\rm g}$ to $\sim 55$~$r_{\rm g}$, or vice versa, towards the end of the observation. Granger lags are a promising way to measure the intrinsic lags, and provide evidence of coronal height variability within each individual observation.

\end{abstract}

\begin{keywords}
accretion, accretion discs -- galaxies: active -- X-rays: galaxies -- X-rays: individual: IRAS13224--3809
\end{keywords}

\section{Introduction} \label{sec:introduction}

The X-ray reverberation technique allows us to study physics of the accretion flows closest to the event horizon of a black hole. This technique is based on a measurement of the delays between the light-travel time of the coronal X-ray photons that are directly observed and those that back-scattered from the accretion disc \citep{McHardy2007, Fabian2009, Uttley2014, Cackett2021}. Then, the variability in the reflection-dominated band (RDB) lags behind the change in the continuum-dominated band (CDB), with the lag amplitude dependent on the source geometry. 

Thus far, the X-ray reverberation models for Active Galactic Nuclei (AGNs) have been developed, and the lags are estimated from the Fourier-phase difference between the light curves in different energy bands \citep[e.g.,][]{Zoghbi2010, Wilkins2013, Emmanoulopoulos2014, Cackett2014, Chainakun2016, Kara2016, Epitropakis2016, Wilkins2016, Caballero2018, Caballero2020}. Probings of the X-ray reverberation signatures on the power spectral density (PSD) of AGN, as well as through the Fourier technique, have been investigated extensively \citep{Emmanoulopoulos2016, Papadakis2016, Chainakun2019, Chainakun2022a}.

Here, we present a detailed study on the Granger causality test \citep{Granger1969}, which is a hypothesis test used to determine whether one time series, e.g. $x_t$, is a factor and can provide useful information in forecasting another time series, e.g. $y_t$. We can state that $x_t$ Granger-causes $y_t$ (or $x \rightarrow y$) if the $x_t$ is statistically useful to predict the future of $y_t$. Comparing this concept to the reflection framework, when the coronal photons illuminate the disc producing the reflection photons, the CDB light curve should Granger-cause the RDB light curve. In other words, the RDB should lag behind the CDB with a specific Granger-lag value. 

In this work, the implications of the Granger causality test for probing the X-ray reverberation lags are explored. The Granger-lag profiles are theoretically modelled and presented, and finally compared to the Granger lags derived from the RDB and CDB light curves of the X-ray reverberating AGN IRAS~13224--3809. In the literature, \cite{Kara2013} reported the flux-dependent reverberation lags in IRAS~13224--3809. The light curves exhibit a complex X-ray variability revealing a significant strong non-stationarity \citep{Alston2019}. It has been thoroughly studied recently via both spectral and timing analysis \citep[e.g.][]{Parker2017, Jiang2018, Alston2020,Caballero2020, Chainakun2022, Hancock2022, Jiang2022}, which suggested a dynamic of the X-ray corona, e.g. the source height increases with luminosity. This makes IRAS~13224--3809 an excellent target for our investigation.

The remainder of this paper is organized as follows. The data reduction for the IRAS~13224--3809 light curves is given in Section 2. In Section 3, we describe the Granger causality test that involves the Augmented Dickey-Fuller (ADF) technique, differencing technique, and the statistical analysis to significantly determine the Granger lags. In Section 4, we explore the modelled Granger lags and compare them to the traditional characteristic-features of the lag-frequency spectra produced using the delta and top-hat response functions. Then, the modelled Granger-lag profiles are compared to those observed in IRAS~13224--3809. The discussion and conclusion are presented in Section 5.

\section{Data reduction}

\begin{table}
\begin{center}
   \caption{\emph{XMM-Newton} observational data of IRAS~13224--3809 used in this work} 
   \label{tab:xmm_obs}
   \begin{threeparttable}
    \begin{tabular}{lccc}
    \hline
    Obs. ID$^{a}$& Rev. no$^{b}$ & Total GTI (ks)$^{c}$ & $^{d}$Duration (ks)\\
    \hline
0673580101	&	2126	&	34.08	&  126.11 \\
0673580201	&	2127	&	49.26	&  125.10 \\
0673580301	&	2129	&	52.03	&  125.00 \\
0673580401	&	2131	&	85.11	&  127.51 \\
0780561301	&	3038	&	112.35	&  134.49 \\
0780561501	&	3043	&	93.22	&  137.49 \\
0780561601	&	3044	&	97.85	&  137.49 \\
0780561701	&	3045	&	100.13	&  137.49 \\
0792180101	&	3046	&	110.64	&  137.69 \\
0792180201	&	3048	&	110.55	&  137.18 \\
0792180301	&	3049	&	86.44	&  130.66 \\
0792180401	&	3050	&	98.63	&  137.48 \\
0792180501	&	3052	&	102.66	&  131.77 \\
0792180601	&	3053	&	101.50	&  132.68 \\
     \hline
     \end{tabular}
    \begin{tablenotes}
    \textit{Note.} $^{a}${\it XMM-Newton} observational ID and $^{b}$revolution number of the data. $^{c}$The total value of GTI within $^{d}$the exposure duration, after performing the data cleaning (see text).   
    \end{tablenotes}
    \end{threeparttable}
    \end{center}
\end{table}

In this work, we applied the Granger causality test to determine the reverberation lags of IRAS~13224--3809, as previously studied in the literature (see Section~\ref{sec:introduction}). All data used in this work was observed by {\it XMM-Newton} observatory \citep{Jansen2001} and obtained from the observatory science archive\footnote{\url{http://nxsa.esac.esa.int/}}; these are tabulated in Table~\ref{tab:xmm_obs}. Here, we selected only the pn observations to obtain high time resolution and high signal-to-noise data. We reduced all observational data using Science Analysis Software (SAS) version 19.1.0 with the latest version of the calibration files.\footnote{\url{https://www.cosmos.esa.int/web/xmm-newton/download-and-install-sas}} The task {\sc epproc} with its default parameters has been applied to generate the calibrated and concatenated EPIC event list file for each observation. We also removed the observational periods, which were highly affected by the background flaring activity, from our data using the SAS task {\sc espfilt} with its default parameters. The remain exposure duration of each observation after we performed the data cleaning is shown in column 4 of Table~\ref{tab:xmm_obs}; whereas the total value of good time interval (GTI) within the exposure duration is shown in column 3 of Table~\ref{tab:xmm_obs}.

We then created the background-subtracted light curves with the time resolution of one second from the events flagged as \#XMMEA\_EP and having PATTERN of $\leqslant$ 4. The source region is the circular area with the radius size of 20 arcsec centred on the source position, while the background region is the circular area of 60 arcsec radius located on the source-free region and also on the same CCD chip as that of the source. Following this method, the soft and hard band light curves are extracted using the energy range of 0.3--1 keV (RDB) and 1.2--5 keV (CDB), respectively; these obtained light curves were then used in the analysis shown in Section~\ref{sec:Comparison with IRAS13224}.

\section{Granger causality analysis}

\subsection{Simulating the AGN light curves}

Let us assume that $a(t)$ is the variation of the X-ray continuum flux. The primary X-ray emission in the continuum-dominated band (CDB) can be written as
\begin{equation}
h(t) = b a(t) \;,
\end{equation}
where $b$ is a normalization factor. The variable emission in the reflection-dominated band (RDB) that contains relatively high reprocessing flux can be written as 
\begin{equation}
s(t) = k a(t) + R\int_0^t a(t^\prime) \psi(t-t^\prime)dt^\prime
 \;, \end{equation} 
where $k$ is a normalization factor, $R$ is the reflection fraction defined as $(\text{reflection flux}) / (\text{continuum flux})$, and $\psi(t)$ is the response function, which is the observed reflection flux as a function of time after the continuum X-ray variations. The realistic response function under the X-ray reverberation framwork can be obtained via the general-relativistic ray-tracing simulations \citep[e.g.][]{Wilkins2013, Cackett2014, Emmanoulopoulos2014, Chainakun2016, Epitropakis2016} by tracing the photons along the Kerr geodesics between the corona, the disc, and the observer.  

Here, the primary variation, $a(t)$, is generated from the {\tt stingray.simulator}, using the shape of the power law spectrum that relates to the red noise processes \citep{Huppenkothen2019}. The normalization factors ($b$ and $k$) can be tied to the flux contribution from the continuum components in each energy band \citep{Emmanoulopoulos2014}. Some authors include the contribution of the reprocessing component in the CDB as well \citep{Chainakun2016, Epitropakis2016}, but  the fraction of reflection found in the CDB is relatively small compared to what found in the RDB, so it is neglected in this work. We then opted to follow the above expressions and set $b=k=1$, so all dilution effects are only controlled by the reflection fraction $R$. The area under the response function is normalized to 1, so that $R=1$ means an equal contribution between the reflection and continuum components.

\subsection{Augmented Dickey-Fuller (ADF) test for stationary}

In order to perform the Granger Causality test, the time series (i.e. light curve) must be stationary, meaning that they must have a constant mean, constant variance, and no seasonal component. Differencing which is a method of transforming a time series by subtracting the previous observation from the current observation is used to remove the components dependent on time (e.g. trend and seasonality). The Augmented Dickey-Fuller (ADF) test, which is based on the unit root test, is used to determine whether a given time series is stationary. The number of unit roots present in the series reveals how many differencing operations are necessary to make the time series stationary.

We use the $p$-value for hypothesis testing to determine whether to reject the null hypothesis. The $p$-value is defined as the probability of obtaining results at least as extreme as what is actually observed if the null hypothesis is true. Generally, when $p<0.05$, the null hypothesis is rejected in favour of the alternative one. The null hypothesis (H$_0$) for the ADF test is arranged such that, if it is not rejected ($p>0.05$), it suggests the time series is not stationary. Alternative hypothesis (H$_{\rm A}$) then indicates that the time series is stationary, which is valid when the null hypothesis is rejected ($p \leq 0.05$).

Here, the differencing and ADF test is performed using the {\tt diff} and {\tt adfuller} modules imported from {\tt statsmodels.tsa.stattools} \citep{Seabold2010}. Note that before executing the Granger causality test, both time series must pass the ADF test to ensure that they are stationary. The differencing method to transform the non-stationary time series into the stationary one is applied, if necessary. The time series may not be stationary after the initial, first-order, differencing in this iterative process. In such circumstances, we must perform a second difference or log transformation until the series is stationary.

\subsection{Granger causality test}

The Granger causality is basically a concept of a hypothetical test in multivariate time series data to verify whether one variable can be used to forecast another variable with a certain lag \citep{Granger1969}. Let us consider two time series in two energy bands under the X-ray reverberation scenario which are the RDB and CDB, referred to as $s_{t}$ and $h_{t}$, respectively. In this case, $h_{t}$ is said to be Granger-cause $s_{t}$ ($h \rightarrow s$) if the past values of $h_{t}$ contain information for predicting $s_{t+1}$ (i.e. $h_{t}$ likely occurs before $s_{t+1}$). The time series $s_{t}$ can then be expressed in a general form of
\begin{equation}
    \begin{split}
        s_{t} = \alpha &  + \beta_{1} s_{t-1} + \beta_{2} s_{t-2} + ... + \beta_{q} s_{t-q} \\
         & + \gamma_{1} h_{t-1} + \gamma_{2} h_{t-2} + ... + \gamma_{q} h_{t-q} ~,   
    \end{split}
    \label{eq:st}
\end{equation}
where $\beta_i$ and $\gamma_i$ are the lag coefficients of $s_{t-i}$ and $h_{t-i}$, respectively. The subscript $i$ runs from 1 to $q$, so there are $2q$ parameters (degrees of freedom) to be estimated. The prediction is done based on the past values of both time series lagged by the order of $q$. 

When $h \rightarrow s$, it implies that $h_{t}$ must precede $s_{t}$ and the lagged values of $h_{t}$ should be significantly related to $s_{t}$. Our null hypothesis (H$_{0}$) for the Granger test is that $h_{t}$ does not Granger-cause $s_{t}$, i.e., $\gamma_{1}=\gamma_{2}=...= \gamma_{p}=0$. An alternative hypothesis (H$_{A}$) then implies that $h_{t}$ does Granger cause $s_{t}$, or that at least one of the lags of $h$ is significant, i.e. at least one of the $\gamma_{i}$ is significantly from zero. Hence the past values of $h_{t}$ have a power to provide more accurate predictions of $s_{t}$ than using the past values of $s_{t}$ alone. 

The Granger-lag selection is about obtaining the best-prediction model. The term of the Granger lag disappears when its coefficient is zero, so the presence of the significant Granger lags is validated by testing if it has significantly non-zero coefficients. In the $h \rightarrow s$ hypothetical test, we then look at the value of the lag number ($q$) that the null hypothesis can be rejected ($p \leq 0.05$). Note that the result of $q$ stands for a test for the joint lag coefficients up to the maximum $q$, not a test for an individual one only. For example, if the H$_{0}$ of the $h \rightarrow s$ test is rejected with $q=10$ ($p \leq 0.05$), this means that at least one of the 10 lag coefficients ($\gamma_{1}, \gamma_{2}, \gamma_{3}, ..., \gamma_{10}$) is significantly different from zero.   

To execute the Granger causality test, we utilize the {\tt grangercausalitytests} module imported from {\tt statsmodels.tsa.stattools} \citep{Seabold2010}. We also perform two-sided Granger test ($h \rightarrow s$ and $s \rightarrow h$) to infer the causality. Since the coronal photons provoke the disc which produces reflection photons that contribute to the RDB with a certain lag, it was expected that the CDB ($h_t$) should Granger-cause (lead) the RDB ($s_t$), but not vice versa. The best way to determine the Granger lags under the reverberation framework is then to select the lag $q$ where the H$_{0}$ of the $h \rightarrow s$ is rejected ($p \leq 0.05$), while the H$_{0}$ of $s \rightarrow h$ is not ($p > 0.05$).

\section{Results and analysis}

\subsection{Modelled Granger-lags with $\delta$-response functions}
\label{sec:GC-Delta}

\begin{figure*}
\centerline{
\includegraphics*[width=0.5\textwidth]{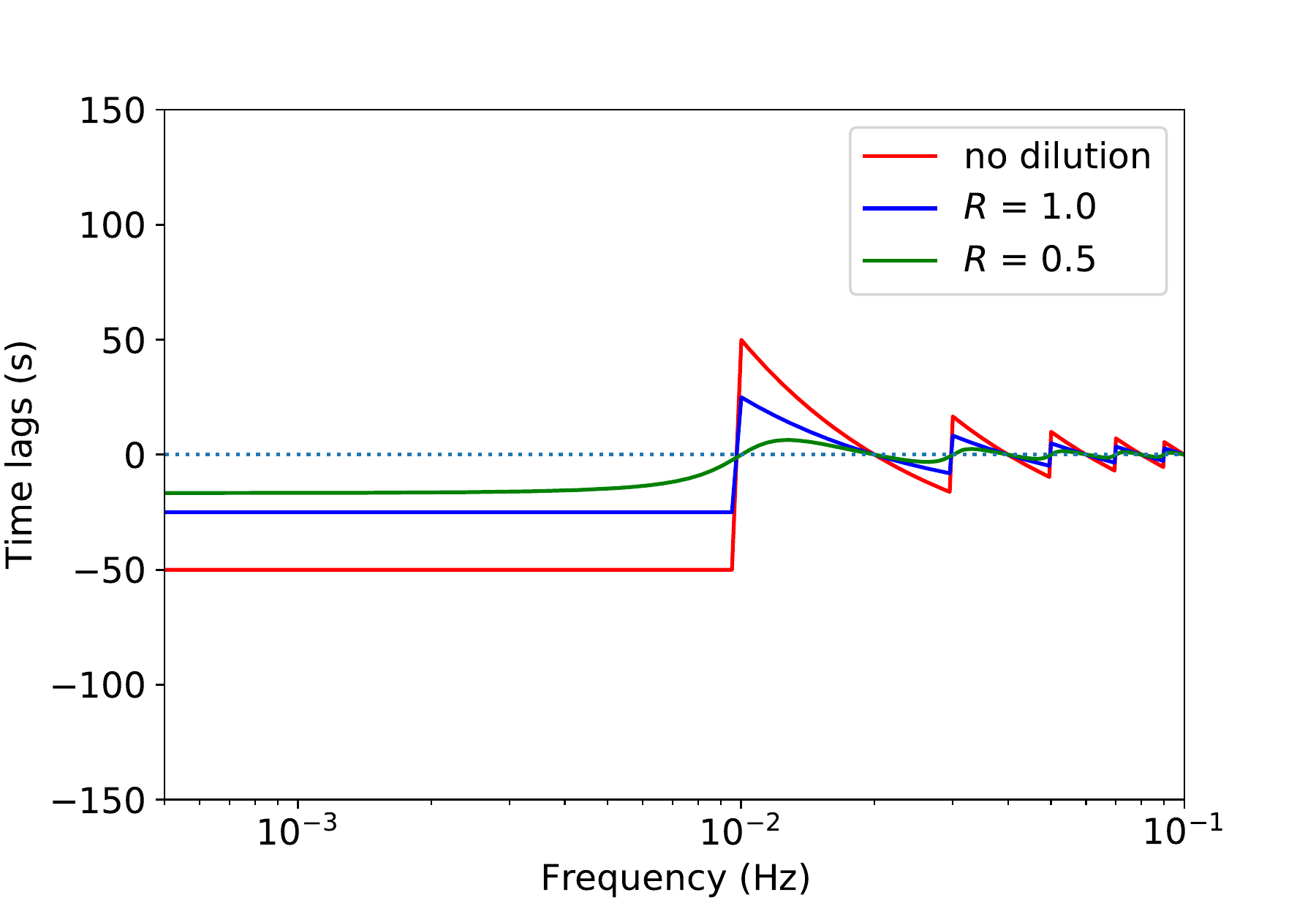}
\put(-210,140){$\delta$-function, $t_{\rm shift}=50$~s}
\hspace{-0.7cm}
\includegraphics*[width=0.5\textwidth]{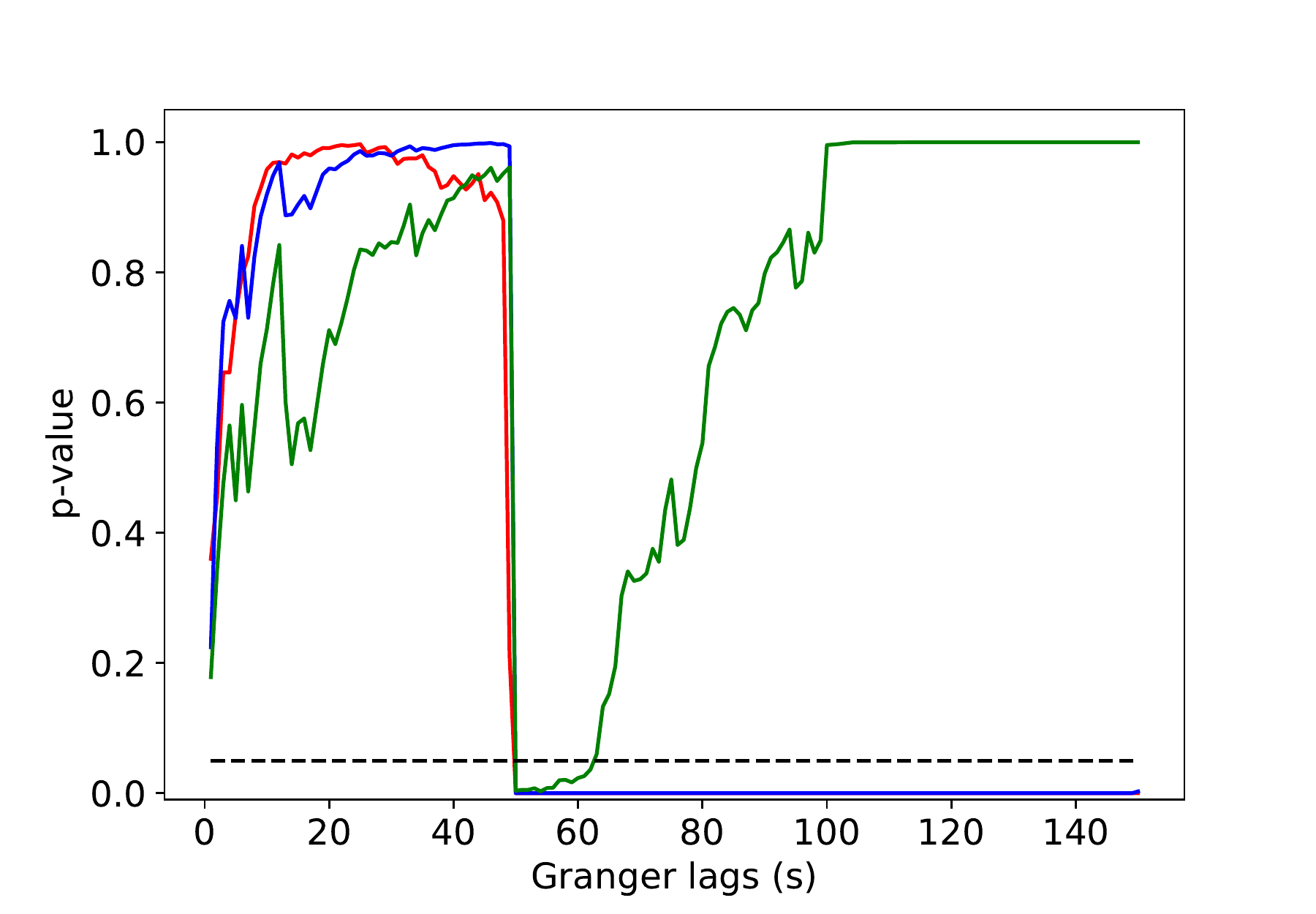}
\vspace{-0.15cm}
}
\centerline{
\includegraphics[width=0.5\textwidth]{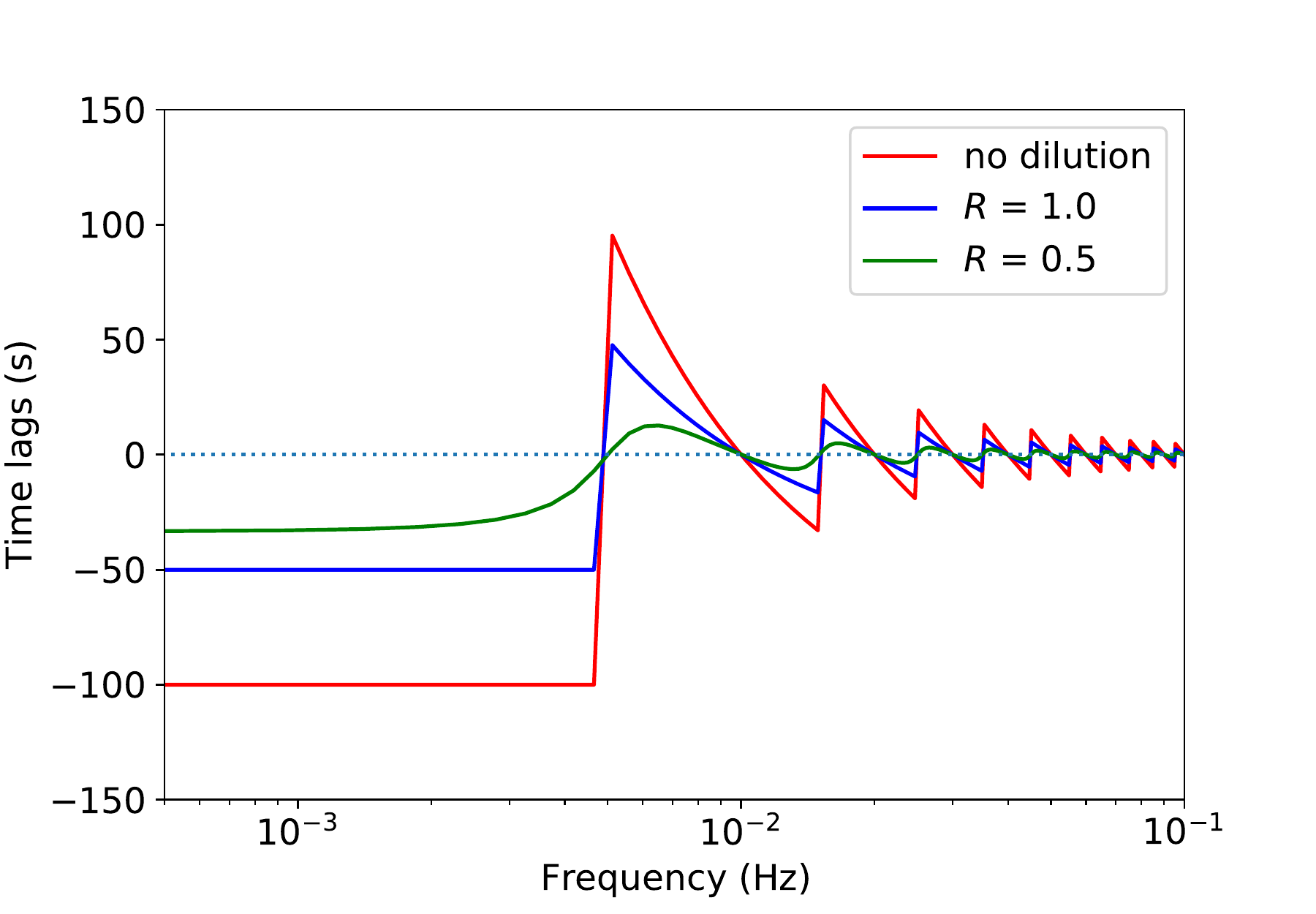}
\put(-210,140){$\delta$-function, $t_{\rm shift}=100$~s}
\hspace{-0.7cm}
\includegraphics[width=0.5\textwidth]{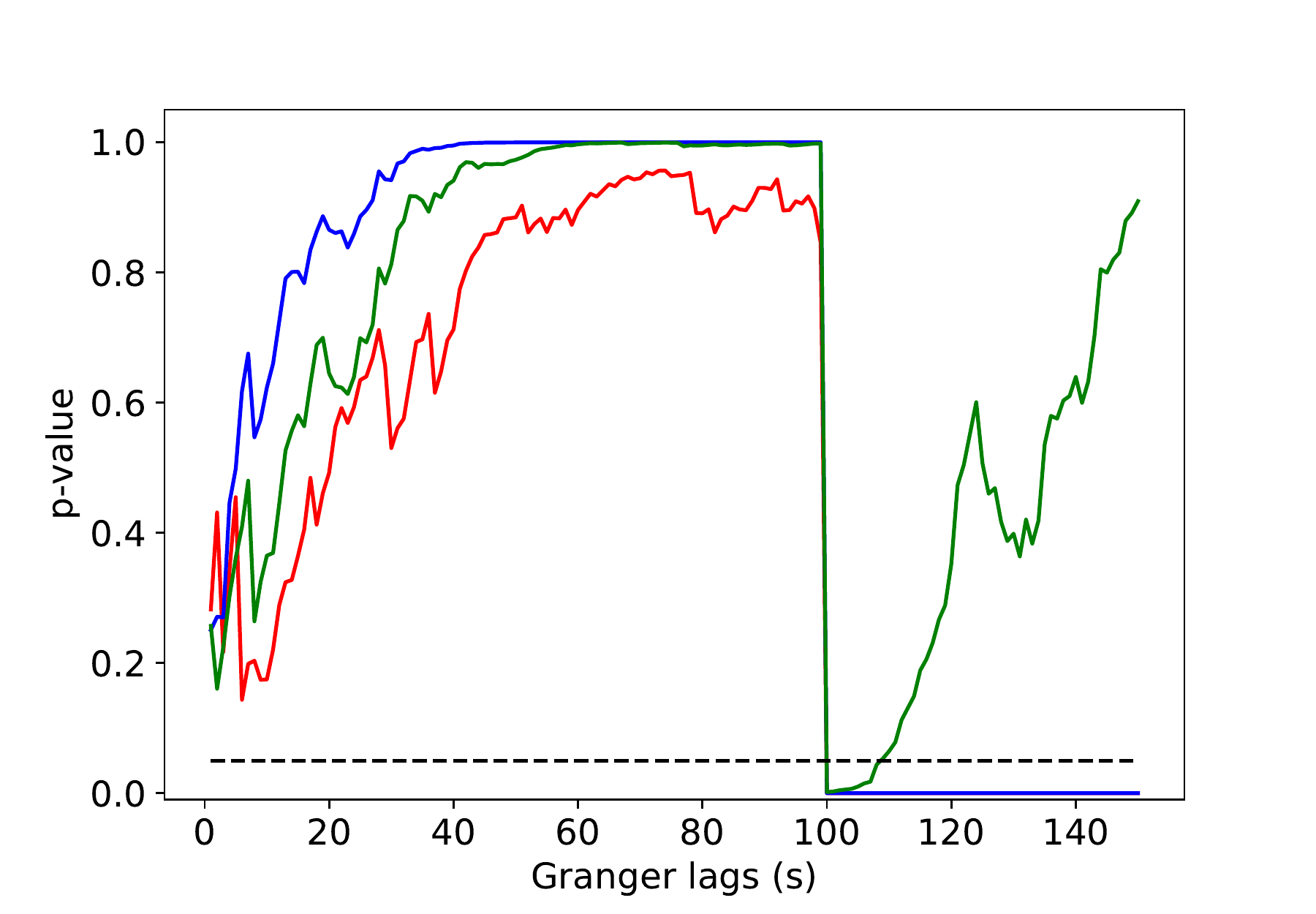}
}
\caption{Left panels: lag-frequency spectra between two light curves related by a time shift of $t_{\rm shift}=$ 50~s and 100~s, when different amounts of dilution are applied. We also show the case when no dilution is applied (i.e. no cross-component contaminated in both energy-band light curves). The time bin of the light curve is $\Delta t=1$~s. Right panels: corresponding Granger-lag profiles for the $h \rightarrow s$ test. The Granger lags start to be significant ($p < 0.05$, below the dashed line) at 50~s (upper panel) and 100~s (lower panel). These Granger-lag values always indicate the intrinsic reverberation lags, regardless of what the dilution amount is applied. The backward tests ($s \rightarrow h$) provides $p > 0.05$ for all ranges of the Granger lags, so their profiles are omitted for clarity. See text for more details.
}
\label{fig-1}
\end{figure*}

\begin{figure}
    \centerline{
        \includegraphics[width=0.5\textwidth]{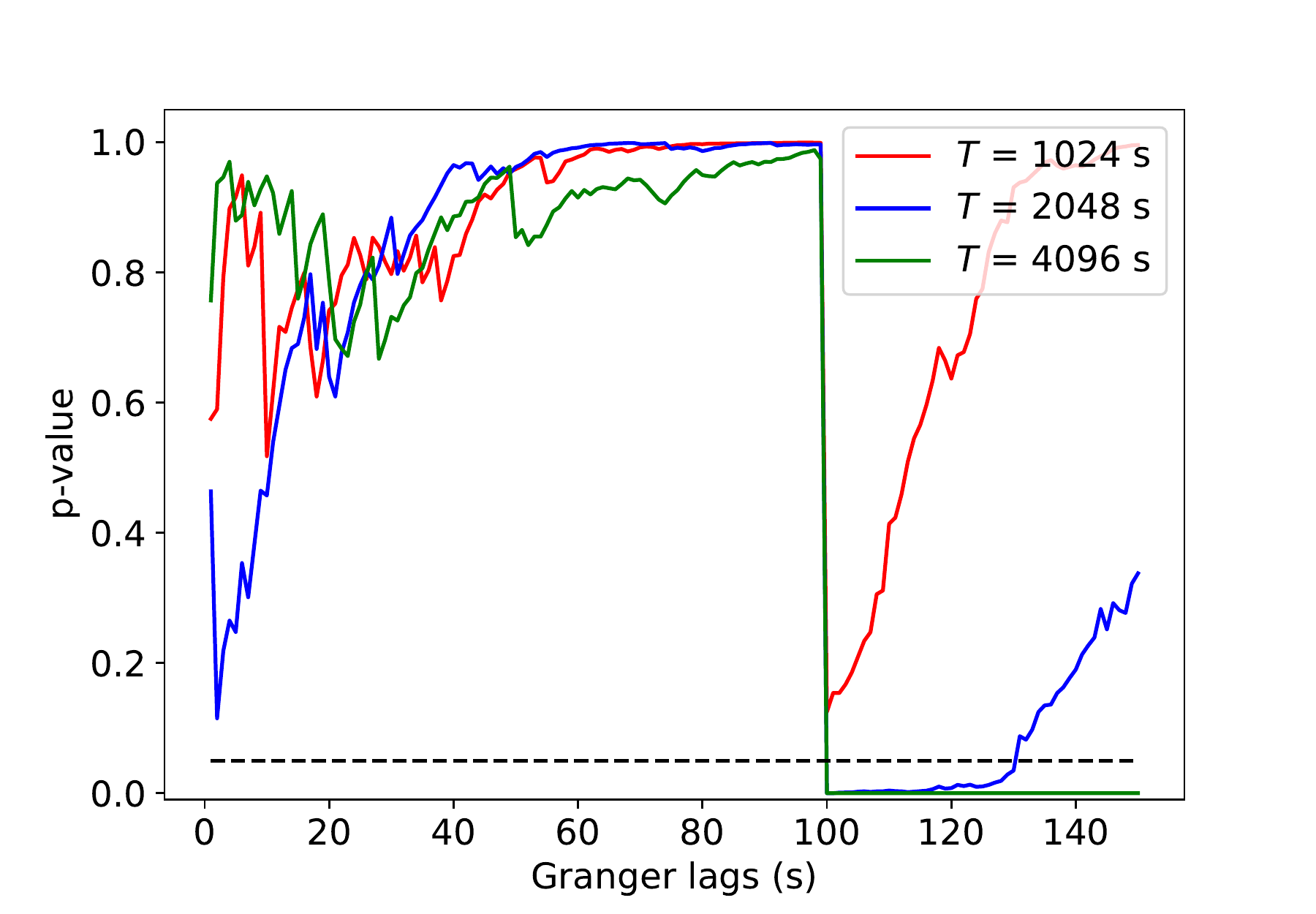}
    }
    \caption{Granger-lag profiles ($h \rightarrow s$) varying with the length of the light curve, $T$, when $R=0.5$, $t_{\rm shift}=100$~s and $\Delta t=1$~s. The Granger lag where it starts to become significant, as before, depicts the amplitude of the intrinsic reverberation lags. However, the light curve with smaller $R$ may require longer series to reveal the significance of the Granger lags. Again, the causality in the opposite direction is also tested, and shown to be insignificant. }
    \label{fig-2}
\end{figure}

First, we investigate the modelled Granger lags when the RDB light curve is produced using the delta-function impulse responses, $\psi(t) = \delta(t-t_{\rm shift}$), which is zero everywhere except at $t = t_{\rm shift}$. For comparison, the traditional frequency-dependent time lags are also calculated such that the negative sign means the variations in the RDB lags those in the CDB. Fig.~\ref{fig-1} shows the lag-frequency spectra and the corresponding Granger-lag profiles varying with $R$ and $t_{\rm shift}$. As expected, we can see that the amplitude of the frequency-dependent lags increases with increasing $R$ and $t_{\rm shift}$. The lags produced by taking into account the dilution effects are always smaller than the intrinsic lags. The $h \rightarrow s$ test, on the other hand, suggests that the Granger lag, where it starts to be significant ($p \leq 0.05$), can be used to indicate the intrinsic X-ray reverberation lag. Note that we also test for the causality in the opposite direction ($s \rightarrow h$), and the lags obtained are not significant at all. This strongly suggests that the CDB Granger-causes the RDB, but the RDB does not Granger-cause the CDB, which is intuitively consistent with the reverberation framework. 

How the Granger-lag profiles vary with the length of the light curve, $T$, is presented in Fig.~\ref{fig-2}. In this case, we fixed $R=0.5$ and $t_{\rm shift}=100$~s. The results show that the minimum Granger lag that is significant is always 100~s, except when $T=1024$~s which is too short for the small $R$. Hence the Granger lags cannot be statistically confirmed (red line in Fig.~\ref{fig-2}). Increasing $T$ indeed allows us to probe the significance of the lags when $R$ is small (i.e., amount of dilution is large). 

Interestingly, we find in some cases that the $p$-values for long Granger lags increase again. This occurs when the reflection fraction is low and when the light curve is short. If the data are good enough, longer Granger lags than the smallest significant value should help predict the light curve in the other band, thus the $p$-values for long Granger lags should be $< 0.05$. When we shorten the light curves or decrease the reflection fraction, the data quality in terms of analyzing the lags are lower. The Granger lags, even if they exist, then become less significant, and the $p$-values start to increase (less significant) from larger to smaller Granger-lag values (e.g. Fig.~\ref{fig-2}).

\subsection{Modelled Granger-lags with top-hat response functions}
\label{sec:GC-TH}

Now, the modelled Granger lags when the RDB light curve is generated using the top-hat (TH) response functions are investigated. The TH function is defined using the parameters that control the centroid, $\tau$, and the width, $t_{\rm w}$. The area under the TH function is normalized to 1. 

\begin{figure}
    \centerline{
        \includegraphics[width=0.5\textwidth]{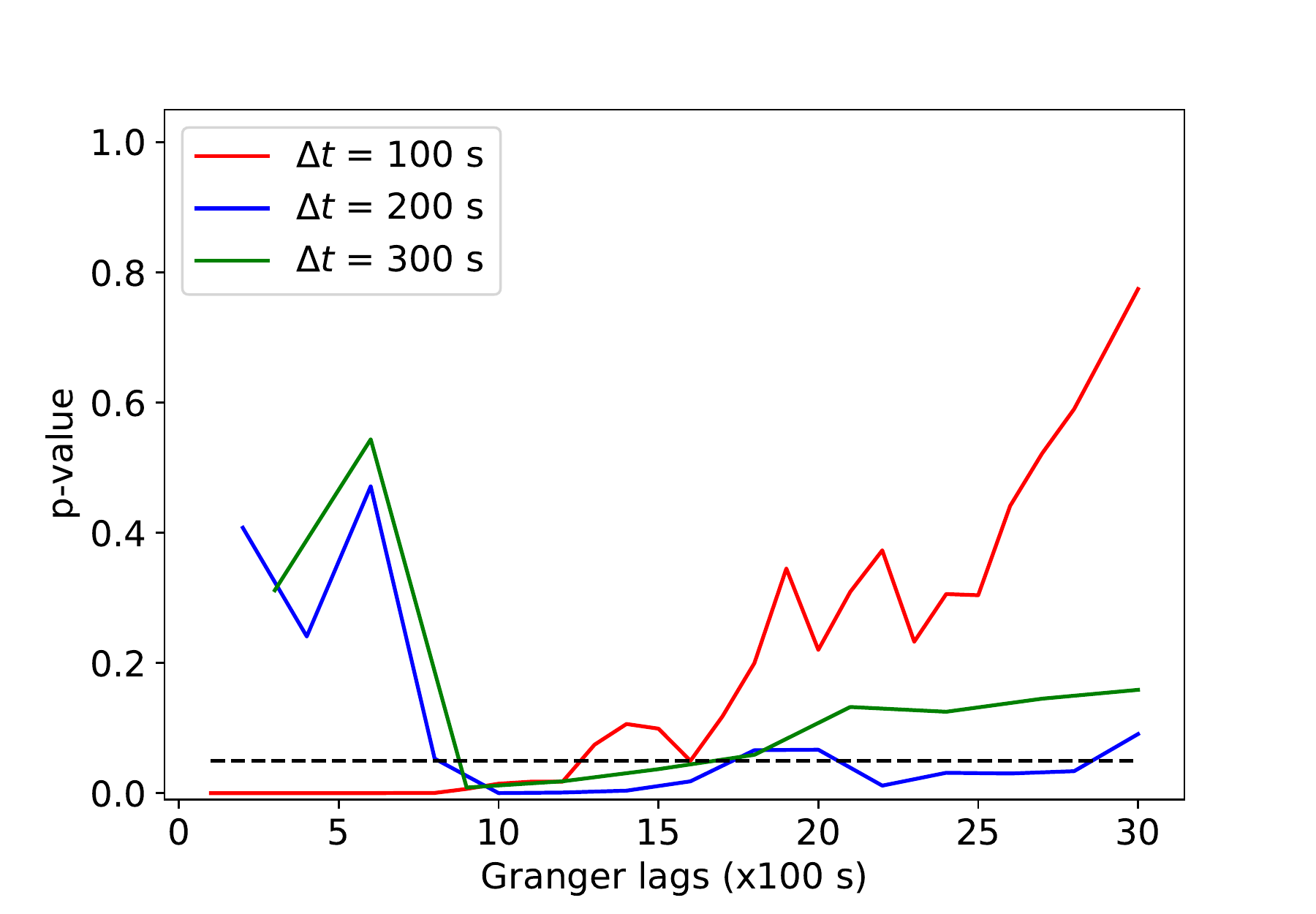}
        \put(-145,144){TH, $\tau=1000$~s}
\put(-131,134){$t_{\rm w}=200$~s}

    }
     \vspace{-0.15cm}
    \centerline{
        \includegraphics[width=0.5\textwidth]{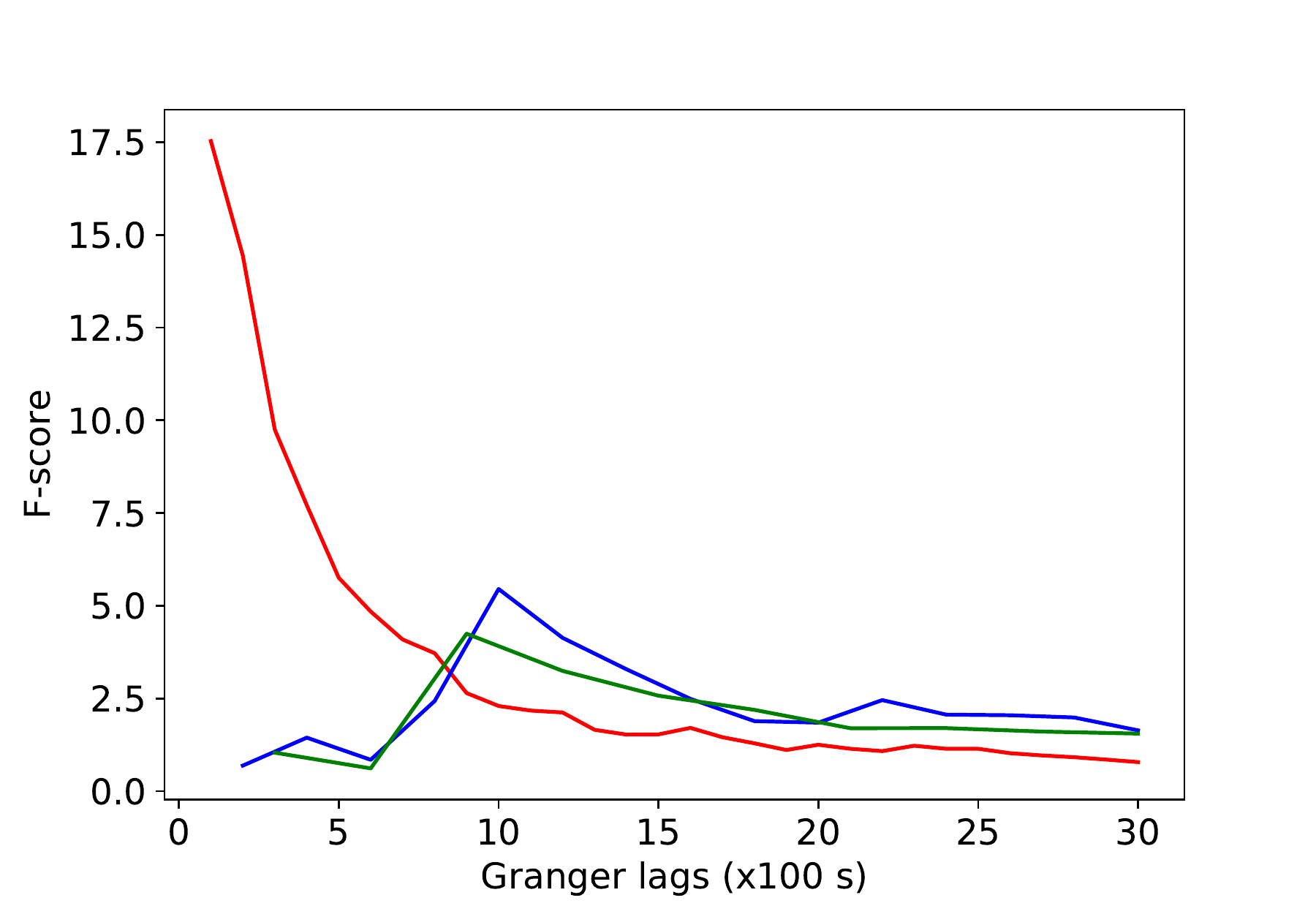}
    }
    \caption{Granger-lag profiles ($h \rightarrow s$) plotted in terms of the $p$-value (top panel) and F-score (bottom panel) when the size of the time bin, $\Delta t$, is varied. The response function is the top-hat with the centroid of $\tau=1000$~s and the width of $t_{\rm w}=200$~s. The reflection fraction is fixed at $R=1$. For clarity, the associating profiles for the causality test in the opposite direction are not shown since the obtained lags are not statistically significant in this direction.}
    \label{fig-3}
\end{figure}

Fig.~\ref{fig-3} (top panel) shows the Granger-lag profiles varying with the size of the time bin $\Delta t$, when the TH function is fixed with $\tau=1000$~s and $t_{\rm w}=200$~s. If $\Delta t < t_{\rm w}$ (red line in Fig.~\ref{fig-3}), the time bin is rather short so that each bin cannot cover the whole response time-interval. Then, the short lag signal which is significant at the lowest time bins can be dominated by the dilution, hence we can still predict some of the soft band from the hard band even at zero lag. Therefore, when the response function is broad, the $\Delta t$ must be chosen to be at least equal or larger than the width of the response function ($t_{\rm w}$), ensuring that the characteristic profiles, as in the case of the $\delta$-response function, can still be comparable. If $\Delta t \gtrsim t_{\rm w}$, the minimum Granger lags at $p \leq 0.05$ can be interpreted as the average intrinsic reverberation lags whose amplitude equals $\tau$. However, we start to see the error in determining these Granger lags. This is due to the binning effects that, at the moment, the lag number in the causality test must be an integer stepping through the time bin size $\Delta t$. The lag error in this case can then be approximated to be $\pm \Delta t$. For example, if we test for the Granger lags of 100, 200, 300, ..., 1000~s (the time bin is 100~s) and find the significant lag of 500~s, the obtained lag with an estimated error would be $500 \pm 100$~s. We also explore the variation and uncertainty of Granger lag measurements induced by random noise or uncorrelated variability in the light curves (see Appendix~\ref{sec:a1}).

To further analyse the Granger lags, we employ the F-score which represents an accuracy result from the measures of fit via the sum of squared residuals (SSR) based F-test. It tests the significance of the coefficients by comparing the variance of the data, and the null hypothesis is rejected when the F-score is large. Fig.~\ref{fig-3} (bottom panel) shows the Granger-lag profiles plotted in term of the F-score. It can be seen that, when $\Delta t \gtrsim t_{\rm w}$, the peak in the F-score profile can be used to identify the reverberation-lag amplitude $\tau$. Due to an uncertainty in determining the bin size, especially in real applications, it might be better to use both $p$-value and F-score profiles to probe the reverberation lags. 

\begin{figure*}
\centerline{
\includegraphics*[width=0.5\textwidth]{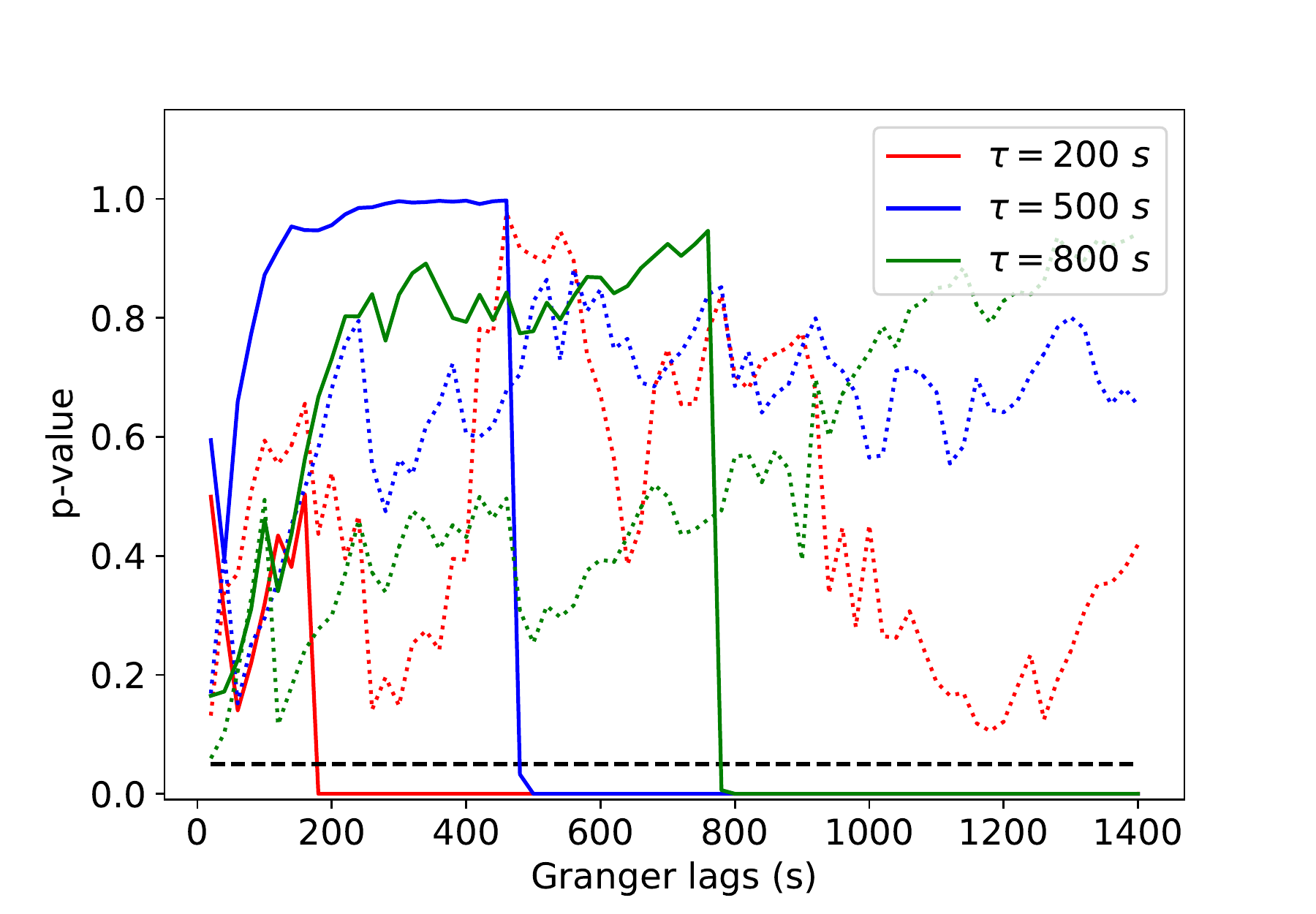}
\put(-187,143){TH, $t_{\rm w}=20$~s, $\Delta t=20$~s}
\hspace{-0.7cm}
\includegraphics*[width=0.5\textwidth]{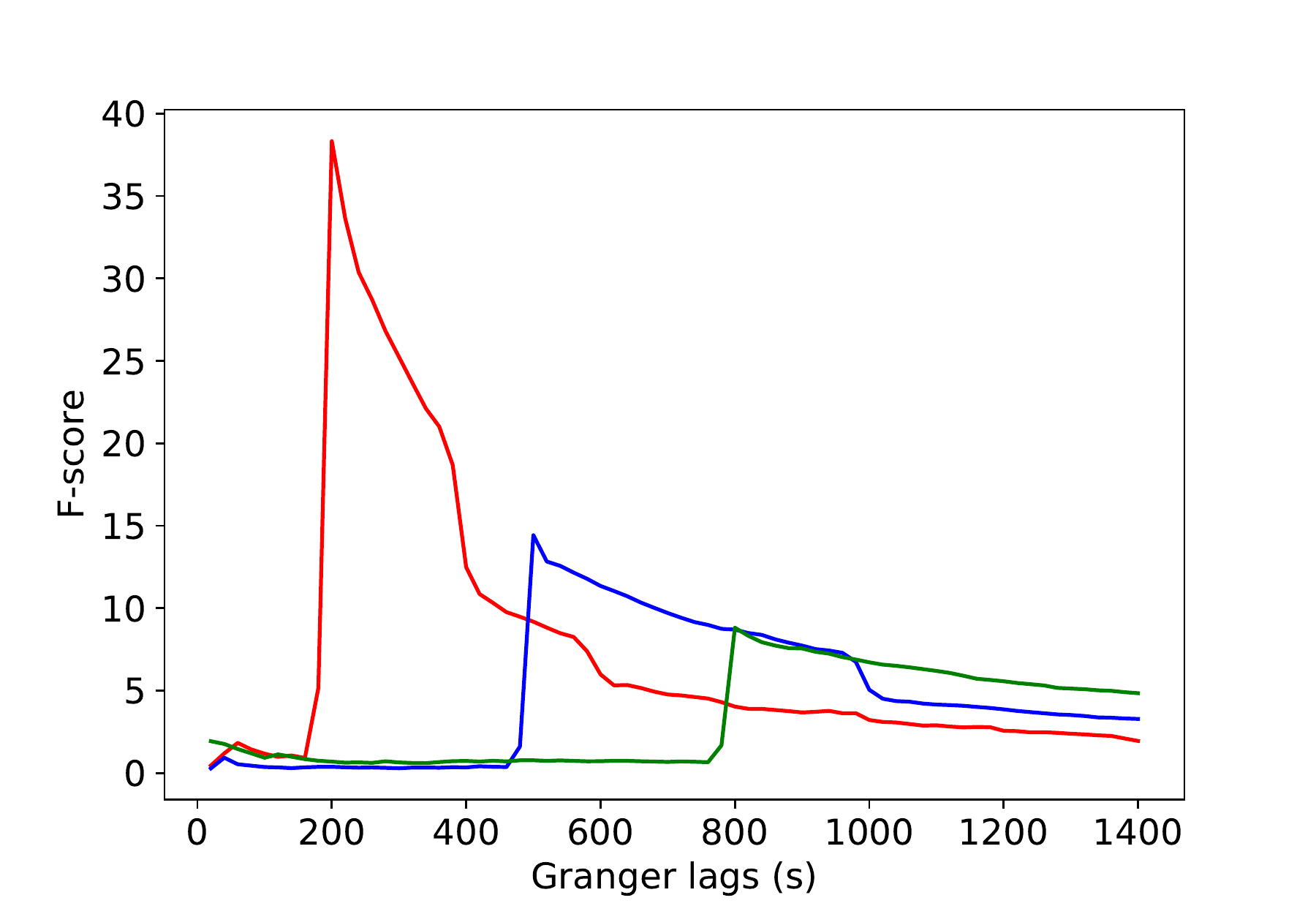}
\vspace{-0.1cm}
}
\centerline{
\includegraphics[width=0.5\textwidth]{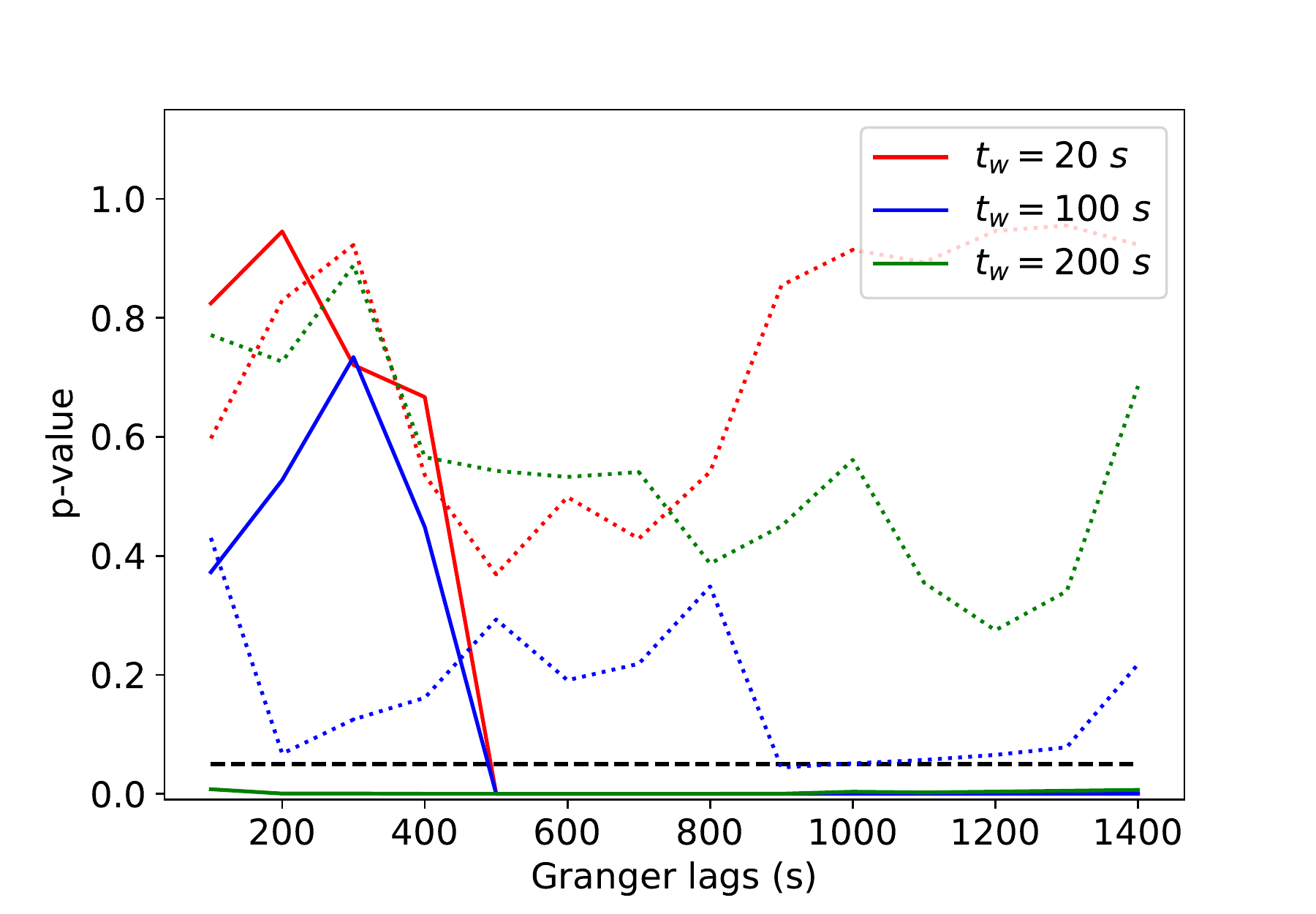}
\put(-195,143){TH, $\tau=500$~s, $\Delta t=100$~s}
\hspace{-0.7cm}
\includegraphics[width=0.5\textwidth]{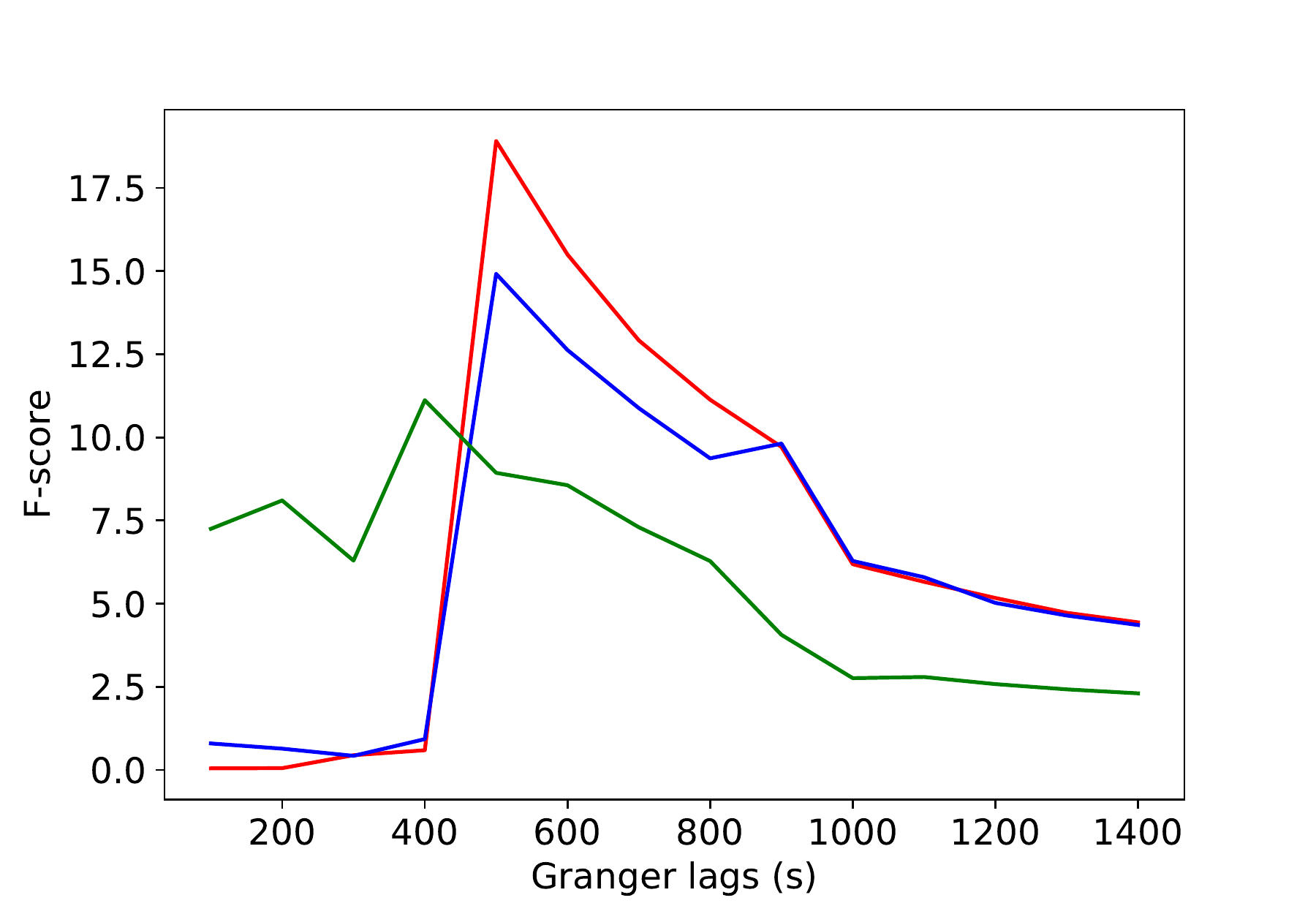}
}
\caption{Top panels: Dependence of the Granger-lag profiles ($h \rightarrow s$) on the top-hat response functions with different centroids ($\tau$) when we fix $t_{\rm w} = 20$~s, $\Delta t =20$~s, $T=20$~ks and $R=1$. Here, we also show the associating $p$-value profiles for the opposite causality directions ($s \rightarrow h$) in the dotted lines. The corresponding F-score profiles (only in the case of $h \rightarrow s$) are presented on the right panels. Bottom panels: Same as the top panels, but varying the width of the top-hat response ($t_{\rm w}$) when we fix $\tau = 500$~s, $\Delta t =100$~s, $T=20$~ks and $R=1$. When the data are binned such that the bin size is $\Delta t \gtrsim t_{\rm w}$ (all cases except the green lines in bottom panels), the minimum lags for the $h \rightarrow s$ test at $p \leq 0.05$ can be used, together with the peak in the F-score profile, to identify the average response times of $\tau$.}
\label{fig-4}
\end{figure*}

To draw a general conclusion, how the Granger-lag profiles probing in terms of the $p$-value and the F statistic varies as a function of the centroid and width of the top-hat response is shown in Fig.~\ref{fig-4}. It is clear that when $\Delta t \gtrsim t_{\rm w}$, the reverberation-lag amplitude of $\tau$ can be inferred from the $h \rightarrow s$ test using the minimum Granger lag at $p \leq 0.05$, with the approximate error of the measured lags of $\pm \Delta t$. This can be statistically confirmed by cross-checking with the F-score profiles whereby we should expect to see the peak of the profiles where the minimum Granger lag appeared in the $p$-value profile as well. Note that binning the data such that $\Delta t \gtrsim t_{\rm w}$ may remove information about the width of the response (i.e. the response behaves simply as if it is a $\delta$-response). This is, however, unavoidable with the current approach, otherwise the obtained Granger lags cannot be easily interpreted. Therefore, the obtained Granger-lag values here, while being interpreted as the intrinsic reverberation lags, should be also thought of as the average response times of the reverberation. Note that the effects of the interplay between the hard and soft lags on the measured Granger lags are also presented in Appendix~\ref{sec:a2}.

\subsection{Comparison with IRAS~13224--3809}
\label{sec:Comparison with IRAS13224}

Before proceeding to analyse the IRAS~13224--3809 light curves, we list below what we know so far from the modeling part, and what we should consider to interpret the Granger lags as the intrinsic reverberation lags: 
\begin{itemize}
\item The light curves must be stationary (i.e. pass the ADF test) before the Granger test is carried out. If the light curves are not stationary, the iterative process of differencing must be initiated to transform the non-stationary light curves until they are stationary.
\item Increasing the length of the light curve may increase the possibility to detect the Granger lags especially when the reflection fraction is low.
\item The light curves must be binned using the bin size of $\Delta t \gtrsim t_{\rm w}$.
\item The minimum Granger lag for the $h \rightarrow s$ test at $p \leq 0.05$, if observed, can be interpreted as the intrinsic reverberation lags. This can be further statistically confirmed if the causality test in the opposite direction ($s \rightarrow h$) for that lag amplitude shows $p > 0.05$ (i.e., no significant lags in the inverse direction), and the F-score profile ($h \rightarrow s$) shows the highest peak at that particular lag too.
\item The identified lags should relate to the average response time and should have an approximate error of $\pm \Delta t$ 
\end{itemize}

In the lamp-post geometry, the realistic response function is broad but not so with the top-hat one; therefore, the width of the response, $t_{\rm w}$, cannot be straightforwardly determined. We, however, expect the response to decay faster at a longer time, so the majority of the response is produced by the reflection from the inner disc regime. At the least, the bin-size used should cover the width of the response before the decay takes place. Although the exact shape of the response function is not known in advance, we can estimate the range of the suitable bin size based on the width of the reverberation response functions from the literature \citep[e.g.][]{Cackett2014, Emmanoulopoulos2014}. Assuming the black hole mass of $2 \times 10^{6} M_\odot$ \citep{Alston2020}, the appropriate time binning for IRAS~13224--3809 is $\sim50$--200~s. We select to bin the data into 20--200~s time bin. The high end of this time-bin choice then should be large enough.

The intrinsic reverberation lags of IRAS~13224--3809 is unknown. We chose to explore the possibility of the Granger lags to probe the intrinsic lags up to $\sim 3000$~s. Note that the light curve segment should be as long as possible to compromise with the uncertainty in the dilution effects that are not constrained here. We then start with using the full light curve for each IRAS~13224--3809 observation.

\begin{figure*}
    \centerline{
        \includegraphics[width=1\textwidth]{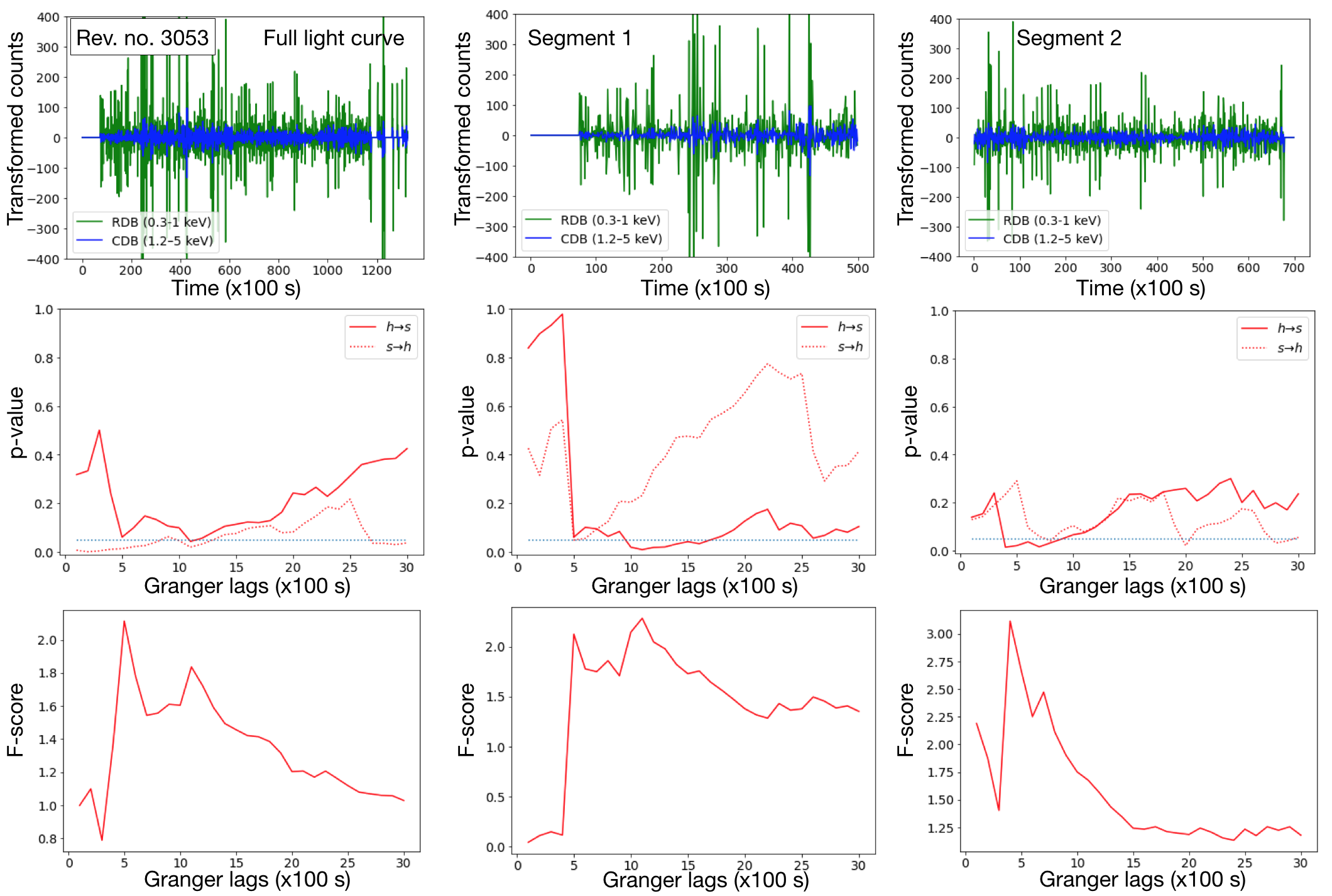}
    }
    \caption{The light curves ($\Delta t= 100$~s) in the soft and hard energy bands of IRAS~13224--3809 in rev. no. 3053 after being transformed to the stationary time series (top-row panels), and the corresponding Granger-lag profiles probing in terms of the $p$-value (middle-row panels) and the F-score (bottom-row panels). The first-, second-, and third-column panels correspond to the analysis using the full light curve, the individual segment no. 1, and the individual segment no. 2, respectively. The red solid lines correspond to the $h \rightarrow s$ Granger test, while the red dotted lines represent the results from the causality test in the opposite direction. The Granger lags that can be interpreted as the intrinsic reverberation lags are found to be statistically significant with the lag amplitude of $\sim 1100$~s and $\sim 400$~s in the segment nos. 1 and 2, respectively. 
    }
    \label{fig-5}
\end{figure*}

An example of the Granger analysis for rev. no. 3053 is shown in Fig.~\ref{fig-5}, where the bin size is fixed at $\Delta t= 100$~s. Initially, the light curve is transformed to be stationary. However, using the full light curve ($\sim 125$~ks), we cannot see the significant Granger-lag signatures that can be interpreted as the reverberation lags (Fig.~\ref{fig-5}, left panels). The light curve is then divided into 2 segments with equal duration, before they are adjusted individually to be $\sim50$~ks and $\sim70$~ks. In this way, we start to see the statistically significant Granger lags that can imply the intrinsic reverberation delays of $\sim 1100$~s and $\sim 400$~s in segments 1 and 2, respectively (Fig.~\ref{fig-5}, second and third columns). In fact, we see the hint of the intrinsic lag-amplitude of either $\sim 400$~s or $\sim 1100$~s using the full light curve as well, but they both become relatively more significant when we consider two segments individually.

\begin{figure}
    \centerline{
        \includegraphics[width=0.5\textwidth]{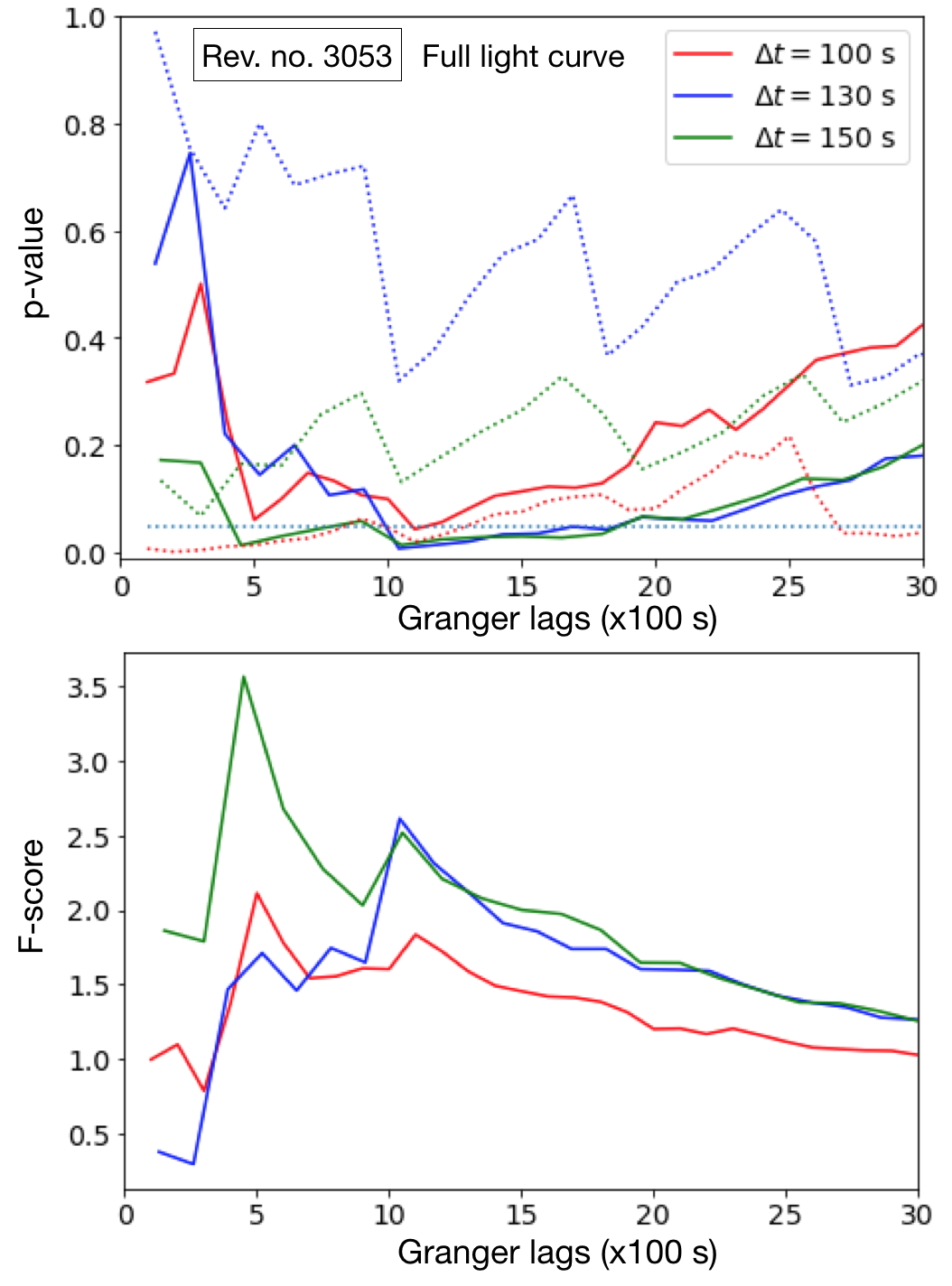}
    }
    \caption{Granger-lag profiles ($h \rightarrow s$) for the full light curve of IRAS~13224--3809 (rev. no. 3053) plotted in terms of the $p$-value (top panel) and F-score (bottom panel) when $\Delta t$ is varied. The profiles for the test in the opposite direction are also shown in the dotted lines. The Granger lags of $\sim 1100$~s and $\sim 400$~s turn out to be significant when $\Delta t=130$~s and 150~s, respectively, consistent with the lags seen in the individual segments shown in Fig.~\ref{fig-5}.
    }
    \label{fig-6}
\end{figure}

The effects of the bin size on the Granger lags for the full light curve in rev. no. 3053 are illustrated, as an example, in Fig.~\ref{fig-6}. It can be seen that the significant Granger lags cannot be probed using the bin size of $\Delta t=100$~s, but when $\Delta t$ is adjusted to be 130~s and 150~s, the Granger lags of $\sim 1100$~s and $\sim 400$~s appear to be significant. Interestingly, these lag values are consistent with the lags seen in the individual segments 1 and 2 in Fig.~\ref{fig-5}. This illustrated that the choice of binning does alter the significance of the lags, and using the full light curve with varying $\Delta t$ should also be able to probe the significant lag values seen in the small segments.

Therefore, for each observation, we use its full light curve and vary $\Delta t$ to be between 20--200~s, in the step of 10~s, and record all significant Granger lags that can be robustly interpreted as the intrinsic reverberation lags. Note that, for each $\Delta t$, the interpretable lag must be the minimum lag value that starts to be significant with $p \leq 0.05$ for the $h \rightarrow s$ test, but $p > 0.05$ for $s \rightarrow h$ test, and the corresponding F-score profile is required to show the highest peak at this lag amplitude. All of these criteria are motivated by our theoretical modelling results (Sections \ref{sec:GC-Delta} and \ref{sec:GC-TH}). 

\begin{figure}
    \centerline{
        \includegraphics[width=0.5\textwidth]{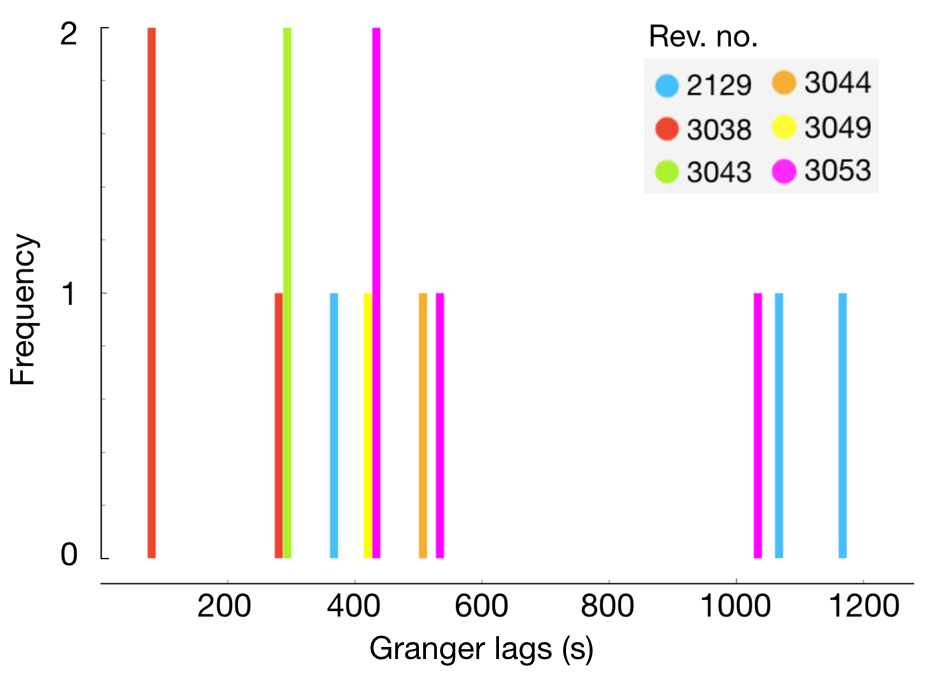}
    }
    \caption{Distribution of the observed Granger lags when using the full light curves. The lags can be identified in 6 observations. While the majority of the Granger lags are $ < 500$~s, there is a gap between the small lags of $< 500$~s and the large lags of $> 1000~s$ in rev. nos. 2129 and 3053.
    }
    \label{fig-sum}
\end{figure}

Fig.~\ref{fig-sum} shows the distribution of the significant Granger lags, which are detected in 6 observations including rev. nos. 2129, 3038, 3043, 3044, 3049 and 3053. It is clear that the majority of the intrinsic lags are $ < 500$~s. However, multiple lags, with either the same or the different values, can also be seen within each individual observation when probed using different $\Delta t$. In rev. no. 3038, the lags of $\sim 100$ and 280~s are found which are not much different; whereas, in rev. nos. 2129 and 3053, the lags are much separate (e.g., lag variability is between $\sim 400$ and 1100~s in rev. no. 3053). 

\begin{figure*}
    \centerline{
        \includegraphics[width=1\textwidth]{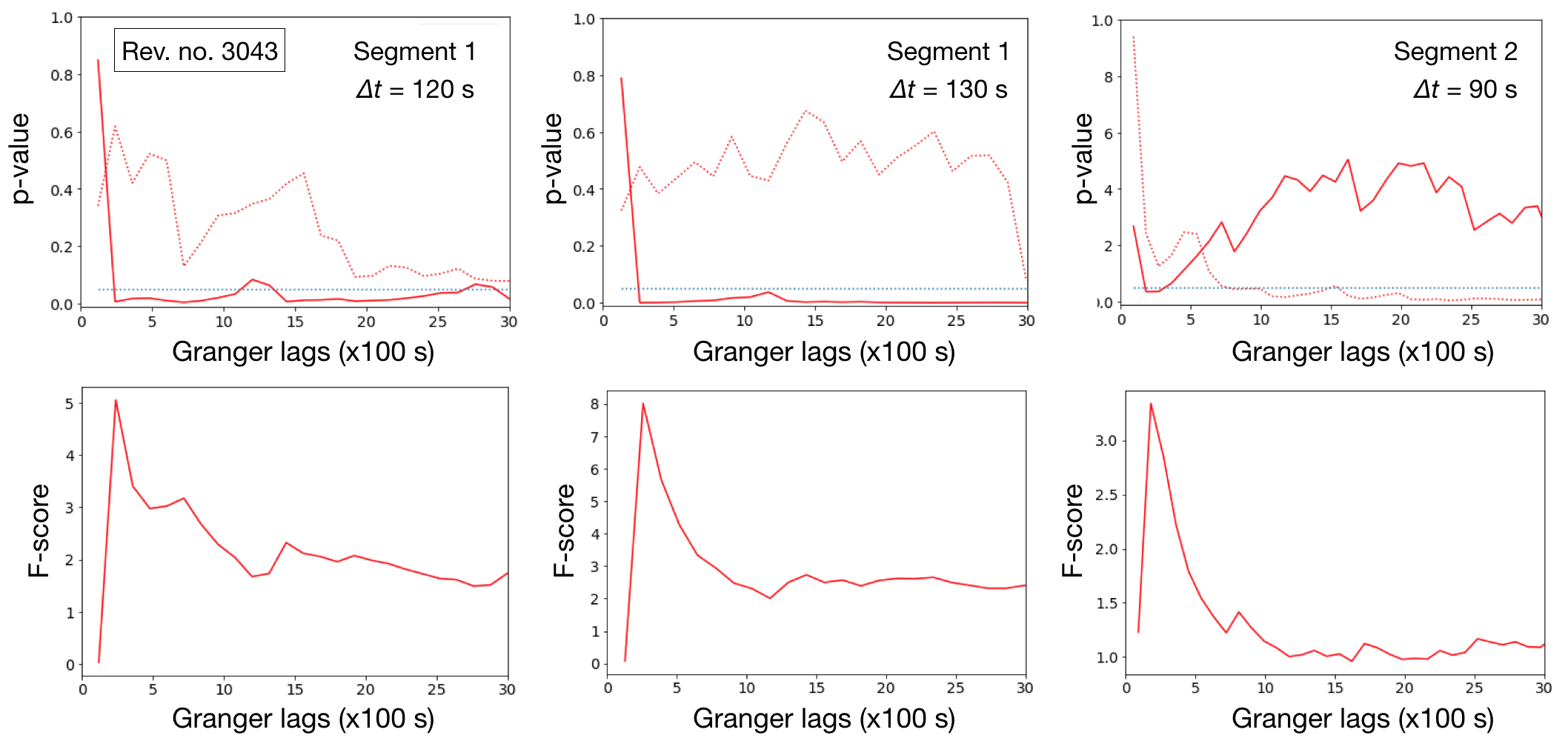}
    }
    \caption{Examples of the Granger-lag profiles ($h \rightarrow s$) for the partial light curves (segments 1 and 2) in rev. no. 3043 probing in terms of the $p$-value (top panels) and the F-score (bottom panels). The profiles for the opposite direction test are shown in the dotted lines. Segments 1 and 2 have approximately the same length, covering approximately the first and second half of the full observation, respectively. 
    }
    \label{fig-3043}
\end{figure*}

\begin{figure*}
    \centerline{
        \includegraphics[width=1\textwidth]{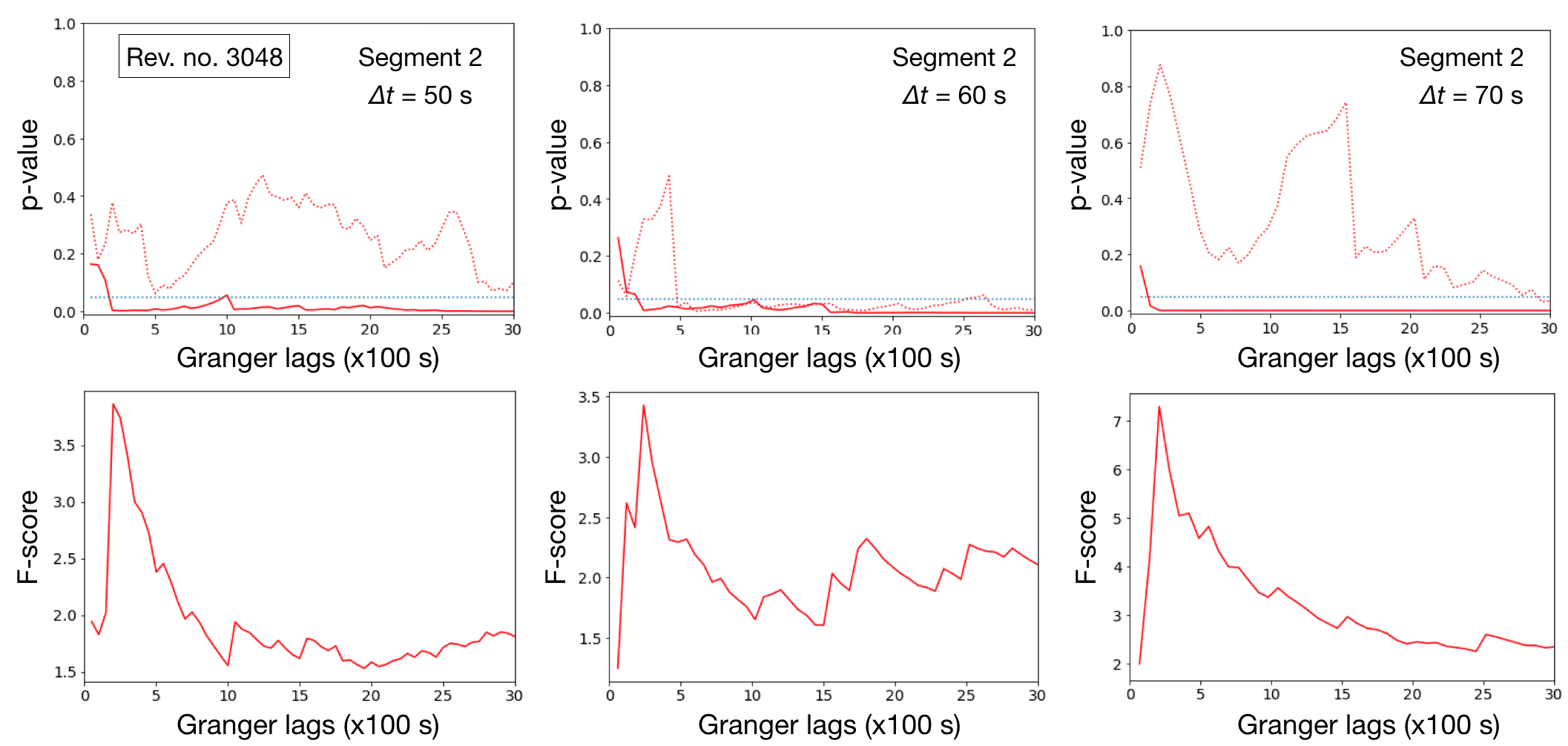}
    }
    \caption{Same as in Fig.~\ref{fig-3043}, but for rev. no. 3048 in which only the lags in segment 2 are found.
    }
    \label{fig-3048}
\end{figure*}

Now, we perform more tests using partial segments of the light curves. To avoid large time-consuming and maintain consistency, each individual light curve is divided equally into two segments whose length is further adjusted, if necessary, by allowing only a small overlapping region between them. Examples of the Granger-lags detected using different $\Delta t$ and in different segments for rev. no. 3043 are presented in Fig.~\ref{fig-3043}. The lags of $\sim 200$--300~s found in both segments are consistent with those found using the full light curve (Fig.~\ref{fig-sum}). Nevertheless, we can detect the significant lags in 6 more observations, including rev. nos. 2126, 2127, 3045, 3048, 3050 and 3052. Therefore, we can detect the Granger lags in 12 observations of IRAS~13224--3809. More lags are probably detected because they are significant only in partial light curves (first half or second half), so they become insignificant when we consider the full length of the observation. This happens in rev. no., e.g., 2126, 2127 and 3048 where the lags are seen only in segment 2 (Fig.~\ref{fig-3048}).

\begin{figure}
    \centerline{
        \includegraphics[width=0.5\textwidth]{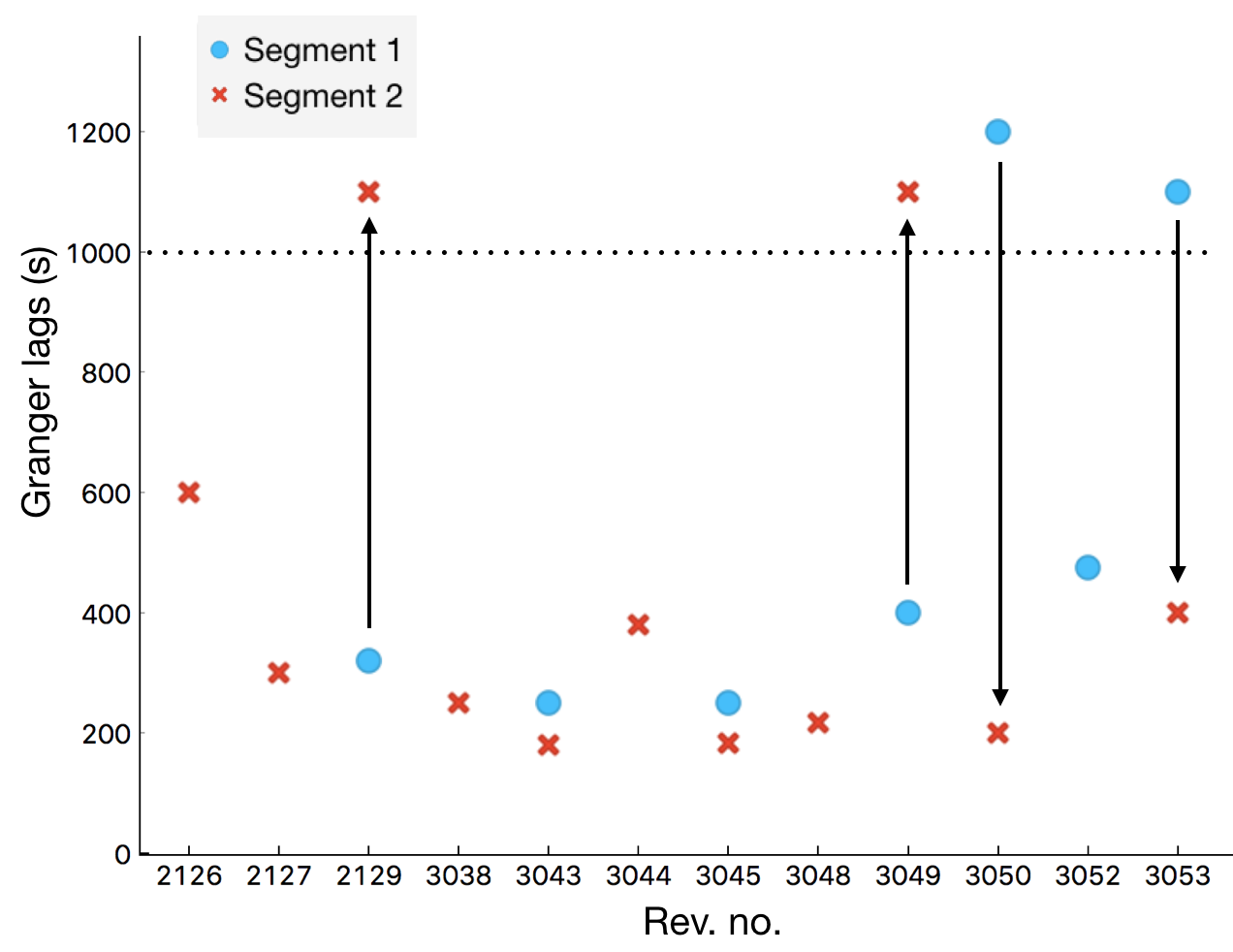}
    }
    \caption{Distribution of the Granger lags in all twelve observations found using partial light curves (segments 1 and 2). The lags in each segment are averaged if more than one lag is found. The gap between the small lags of $< 500$~s and the larger lags of $> 1000$~s is also seen, similar to Fig.~\ref{fig-sum}. The errors for these lags are varying between 50--200~s, estimated from the bin size used to find these lags. A significant increase (rev. nos. 2129 and 3049) and decrease (rev. nos. 3050 and 3053) of the lags towards the end of the observations are clearly seen, which may suggest a height-changing corona in each individual observation.
    }
    \label{fig-final}
\end{figure}

Furthermore, we still see the gap between the small lags of $< 500$~s and the larger lags of $> 1000$~s. Fig.~\ref{fig-final} shows the distribution of the observed lags in all 12 observations using partial light curves (segments 1 and 2). In each segment, the lags are averaged if more than one lag is found. A sudden increase of the lags towards the end of the observation is clearly seen in rev. nos. 2129 and 3049. Contrarily, in rev. nos. 3050 and 3053, the lags are significantly large at the beginning before falling out at the end. A significant increase and decrease of the lags in these observations may suggest the variability of the coronal height within each individual observation. Despite this, we find no clear trend of how the increase or decrease in time lags relates to a change in the light curves (e.g. a change in the count rate or flux). More discovered data are required to relate the changes in time lags in each individual observation to evidence of variability.  

Note that the Granger causality is the statistical test revealing that the lag at a specific value is either significant or insignificant. The more Poisson noise and other noise processes contaminate in the light curve, the more likely that the lag is insignificant. It might not be straightforwards how uncertainties or errors for the Granger lags can be estimated. Here, we use the size of the time bin as the approximate errors of the lags, so all the errors of the lags reported for IRAS~13224--3809 are $\pm (50$--200)~s. See further detail on the uncertainty of the Granger lags due to random noise in Appendix~\ref{sec:a1}.

\section{Discussion and conclusion}

The Granger causality test \citep{Granger1969} is applied to study the relationship between the AGN light curves (CDB and RDB), which are stochastic in nature. The disc response function is broad and the average reverberation delays may be large, making it difficult to estimate significant Granger lags from short observations. To avoid the final light curve being dominated by the artificial data, we select not to fill in the gap of the light curves. Using the full light curve, the IRAS~13224--3809 data in six observations can pass validity and reliability tests based on the $p$-value F statistics. Using the half light curves (segment 1 and 2), the lags in six more observations are found (so twelve in total). This confirms that information from the CDB helps explain the RDB, in agreement with the well-known reflection framework, i.e., while the CDB light curve is dominated by the direct coronal photons, the RDB light curve is dominated by the coronal photons that are being reprocessed via the disc before being observed with a specific time delay \citep{Uttley2014, Cackett2021}. 

In the standard framework, the reverberation and disc propagating-fluctuations produce the lags on different timescales and in different directions, which are the soft lags ($h \rightarrow s$) and hard lags ($s \rightarrow h$), respectively, according to the energy bands considered here. The hard lags from the propagating-fluctuation process should require significantly larger bin sizes to probe, since the width of its response function should be significantly larger. The Granger lags found in the $h \rightarrow s$ direction using the bin size as what we apply for IRAS~13224--3809 is then driven by the reverberation (assuming that no other mechanisms can produce soft lags on comparable timescales). The competing processes such as the propagating-fluctuations should not significantly manifest or dominate the lags in this direction, otherwise we should not be able to detect the significant Granger lags in the $h \rightarrow s$ direction in the first place (see also Appendix~\ref{sec:a2}).

Granger causality does not imply true causality, so it does not require true causal interactions between the CDB and RDB light curves. It is a statistical test showing whether at least one of the coefficients for the lags of one light curve on the other are significant or not. In the traditional way of producing the lags through the Fourier techniques \citep[e.g.][]{Demarco2013, Cackett2014, Chainakun2016, Kara2016, Caballero2018, Caballero2020, Demarco2022}, the contamination between cross-components in both energy bands are jointly affected time lags, and the observed lags are the diluted ones. High density disc model would also produce higher reflection flux at $<2$~keV \citep{Jiang2022}, resulting in higher reflection fraction compared to those obtained using the standard accretion disc \citep{Jiang2018}. Our results show that the Granger lags, if observed, are not affected by the reflection fraction that controls the amount of dilutions (e.g. Fig.~\ref{fig-1}--\ref{fig-4}). Small reflection fraction would make the Granger lags difficult to be probed with statistical significance, but if detected, they are always the intrinsic lags, not the diluted lags.   

Our results provide insight into the characteristics of Granger-lag profiles that link to the intrinsic-lag variation induced by the corona evolution in IRAS~13224--3809, as discussed below.

\subsection{The intrinsic lags and reflection fraction}

In principle, the intrinsic lags can be approximated from the observed lags if the reflection fraction is known. For example, in SWIFT~J2127.4+5654, the Compton hump (RDB) was found to lag behind the continuum (CDB) by $\sim 500$~s, while the reflection fraction of the Compton-hump and continuum band is $\sim 50$ and 10 per cent, respectively \citep{Kara2015}. By working backwards, the observed lags should be $\sim 40$ per cent of the intrinsic lags, hence the intrinsic lags are around 1250~s. However, this is likely model dependent, meaning it is based on a measurement of the reflection fraction that depends on the choice of the reflection models. In the AGN, such as IRAS~13224--3809 where the time-average spectrum is particularly complex \citep[e.g.][]{Fabian2013, Chainakun2016,Jiang2018,Jiang2022}, determining the amount of dilutions and reflection fraction from the spectral analysis is definitely uncertain. The Granger causality test then should provide a more direct way to probe the intrinsic lags. 

Previous literature found the observed, diluted reverberation lags of $\sim 100$~s in IRAS~13224--3809 \citep[e.g.][]{Alston2020, Caballero2020}. The majority of the Granger lags found here are $\sim 200$--500~s, so the diluted lags are $\sim 20$--50 per cent of the intrinsic lags. If the difference between the intrinsic and the observed lags in IRAS~13224--3809 is entirely due to dilution, it will require the spectral dilution factor of $\sim 0.5$--0.8. The strongest dilution factor implied here is around the high end proposed by \cite{Wilkins2013} that the observed time lags can be reduced by up to 75 per cent from the intrinsic lags due to the contamination of the reflection (continuum) flux in the CDB (RDB). For a few observations that also show the large Granger lags of $> 1000$~s, the difference between the intrinsic and observed lags may require the soft corona at large height to also move away from the black hole, producing more directly observed flux which is spectrally soft which can further suppress the intrinsic lags. 

\subsection{The coronal height}

IRAS~13224--3809 contains a black hole mass of $\sim 2 \times 10^{6} M_{\odot}$ \citep{Alston2020}, so a distance of 1~$r_{\rm g}$ corresponds to a light-travel time of 10~s. The intrinsic small lags of $\sim 200$--500~s appeared in almost all observations then correspond to a true light-travel distance of 20--50~$r_{\rm g}$. The black hole spin is likely maximum \citep{Fabian2013, Chiang2015, Jiang2018}, so the inner edge of the disc is very close to the event horizon. For simplicity, let us assume a face-on disc so the coronal height is approximately half of the light-travel distance between the corona and the disc. This places the corona at $\sim 10$--25~$r_{\rm g}$ above the accretion disc, with the uncertainty of $\sim 5 r_{\rm g}$ approximated by an average $\Delta t$ of 100~s. Focusing on six observations where the Granger lags are detected, the range of source height was previously reported to be $\sim 6$--17~$r_{\rm g}$ found by fitting the lag-frequency spectra \citep{Alston2020} and $\sim 3$--16~$r_{\rm g}$ based on the PSD analysis \citep{Chainakun2022}. These heights are comparable to what was implied from our Granger lags.

If the corona changes a lot within an observation, the Granger causality that significantly changes over time may be observed in partial light curves. This is clearly evidenced in some observations where multiple lags are detected within an individual observation. For rev. nos. 3043 and 3045, two different Granger lags are seen in different segments but both are still $< 500$~s (Fig.~\ref{fig-final}). On the other hand, the lags of $\sim 1000$--1100~s are also seen in rev. nos. 2129, 3049, 3050 and 3053. This lag variation puts the corona far beyond the high ends reported in the literature \citep[e.g.][]{Alston2020, Caballero2020, Chainakun2022, Jiang2022}. Perhaps this is evidence of the coronal height variability within these individual observations. For example, in rev. nos. 3050 and 3053, the corona moves closer to the black hole from $\sim 55 r_{\rm g}$ to $\sim 20 r_{\rm g}$ during approximately the first and second half of the observation, producing the lags of $\sim 1100$~s and $\sim 400$~s, respectively. Contrarily, in rev. nos. 2129 and 3049, the corona significantly moves away from the center toward the end of the observation (Fig.~\ref{fig-final}). The variability of the lags are averaged out when we calculate the lags in the traditional way.  

In the X-ray reverberating AGN, the RDB light curve usually has lower amplitude variability than the CDB light curve \citep[e.g.][]{Zoghbi2010}. This is explained by the light bending model (LBM) where the corona changes its height above the disc surface \citep{Miniutti2004}. We then expect the height-changing corona in each particular observation to induce the effects from the LBM. The corona evolution that produces the intrinsic-lag variation seen in an individual observation of IRAS~13224--3809 then much supports the LBM. 

Moreover, with a fixed source height, an additional light-travel time could possibly be obtained if we consider the returning radiation \citep{Wilkins2020}, i.e. by allowing the reflected emission to be returned to the disc, traveling longer distance and producing higher order reflections. However, for our large source height of $\sim 10$--25~$r_{\rm g}$ that is possibly up to $\sim 55 r_{\rm g}$, inferred from the intrinsic lags, the fraction of the returning radiation is small. Therefore, the corona is just slightly more compact even if the effects of the returning radiation is applied. We, however, note that the inferred coronal height here is just a rough approximation. The Granger lags should be compared to the average response times of the response functions and the intrinsic lags in the Kerr spacetime can also be increased by a few $t_{\rm g}$ in the proximity of the black hole. In this case, the relativistic ray-tracing simulation is still required to precisely constrain the source geometry, perhaps by 
mapping the intrinsic lags with the average response times obtained by integrating through the realistic disc response functions that link to each specific coronal height.

\subsection{The future prospects}

While much of the focus is on using either the full or the approximately half light curves, the Granger test may also detect the causality in a more specific time interval within an individual light curve. Probing the causal relationship in such time intervals (i.e. dividing the whole light curves into several partial segments and/or grouping them into similar flux states) might require much effort in practice, but it is worth investigating in the future. How this method could be applied for revealing the thermal reverberation lags in X-ray binaries \citep{Uttley2011, Demarco2015, Mahmoud2019, Chainakun2021}, or to detect the lags in different reflection scenarios such as the tidal disruption events \citep{Kara2016b} and multiple scattering from the outflowing wind \citep{Luangtip2021} is also the subject for future study. 

Finally, while this work outlines the method for probing the intrinsic reverberation lags through the Granger causality test, the aim is not to improve the Granger technique itself. Therefore, there is still room for improvement in the technical details, such as how to engineer the Granger algorithm for a more robust detection, re-inventing the way we implement it, and using a variable-length sliding window to probe the Granger lags thereby providing a new perspective.

\section*{Acknowledgements}
We thank the anonymous referee for insightful comments and useful suggestions that improved the paper. This work was supported by Office of the Permanent Secretary, Ministry of Higher Education, Science, Research and Innovation (OPS MHESI), Thailand Science Research and Innovation (TSRI), and Suranaree University of Technology (grant No. RGNS~64--118). PC thanks funding support from the NSRF via the Program Management Unit for Human Resources \& Institutional Development, Research and Innovation (grant number B16F640076). NN thanks Suranaree University of Technology (SUT), Thailand Science Research and Innovation (TSRI), and National Science Research and Innovation Fund (NSRF), under project no. 90464.

\section*{Data availability}
The AGN data underlying this article can be accessed from {\it XMM-Newton} Observatory (\url{http://nxsa.esac.esa.int}). The modules for the Granger causality test is adopted from {\tt statsmodels.tsa.stattools} available in \url{https://www.statsmodels.org}. The derived data and developed model underlying this article will be shared on reasonable request to the corresponding author.

\appendix

\section{Effects of random noise on the uncertainty of Granger lags}
\label{sec:a1}
\begin{figure*}
\centerline{
\includegraphics*[width=0.5\textwidth]{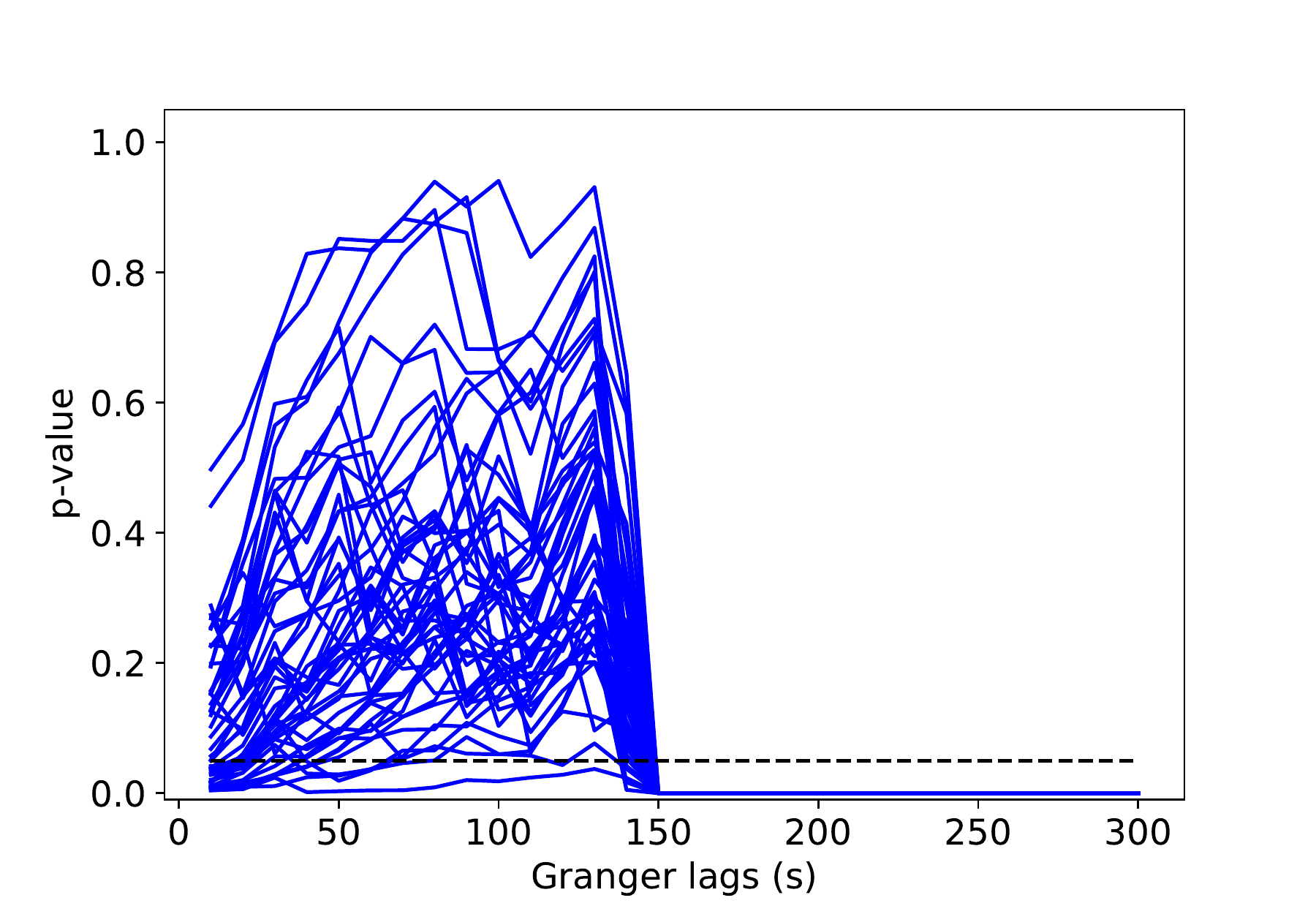}
\put(-117,143){$\sigma=0.1$, $\Delta t=10$~s}
\hspace{-0.7cm}
\includegraphics*[width=0.5\textwidth]{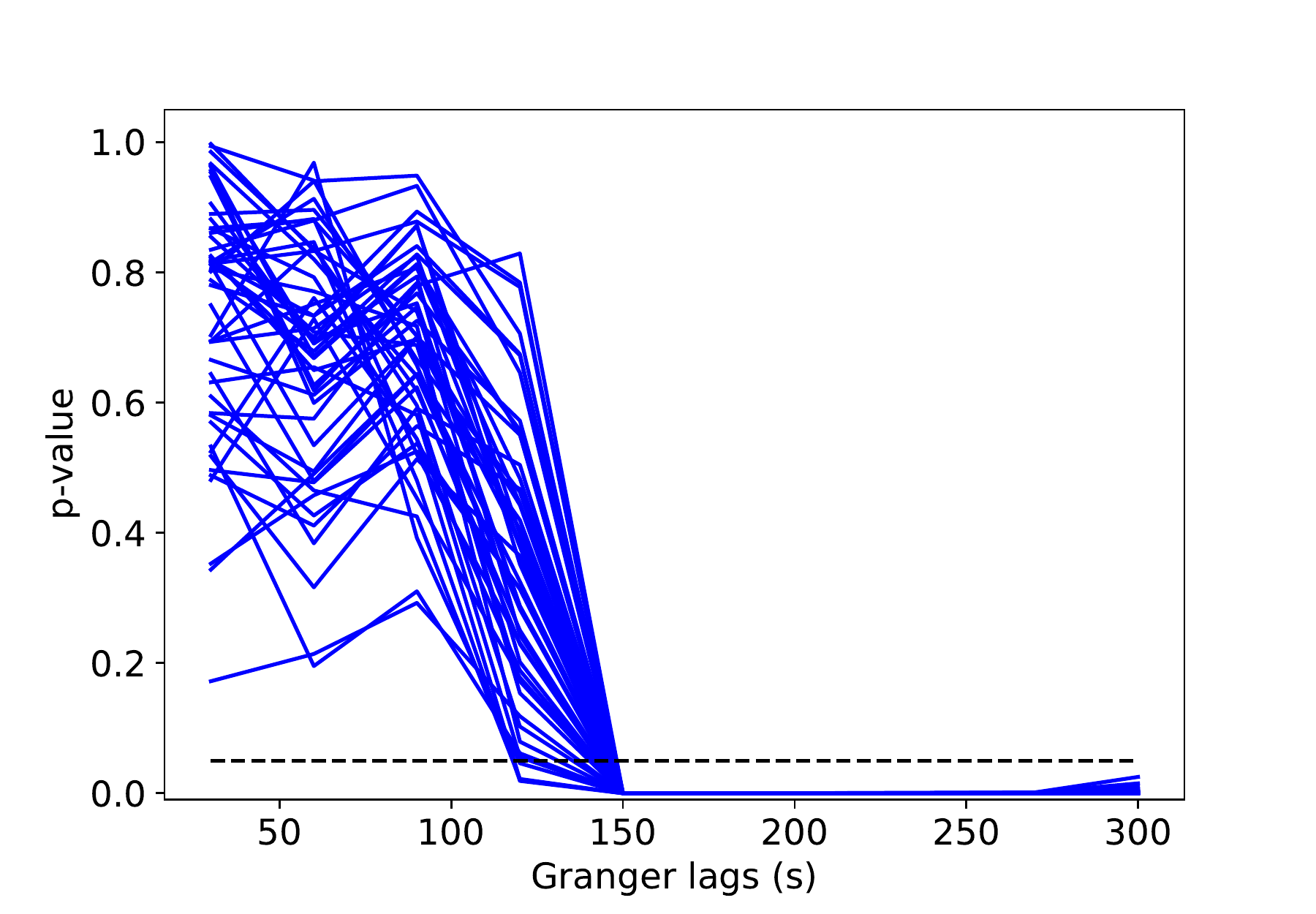}
\put(-117,143){$\sigma=0.1$, $\Delta t=30$~s}
\vspace{-0.1cm}
}
\centerline{
\includegraphics[width=0.5\textwidth]{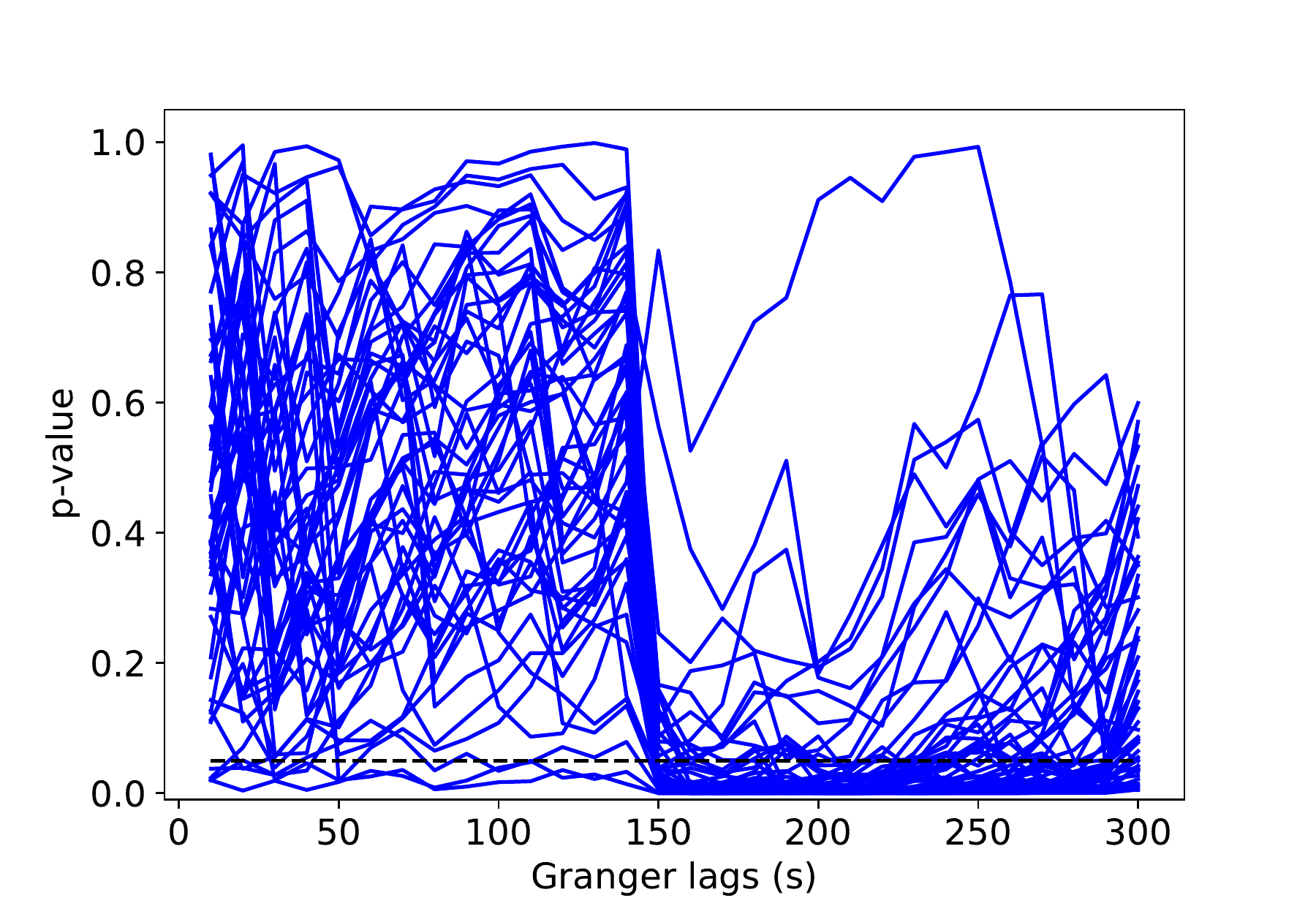}
\put(-117,143){$\sigma=0.3$, $\Delta t=10$~s}
\hspace{-0.7cm}
\includegraphics[width=0.5\textwidth]{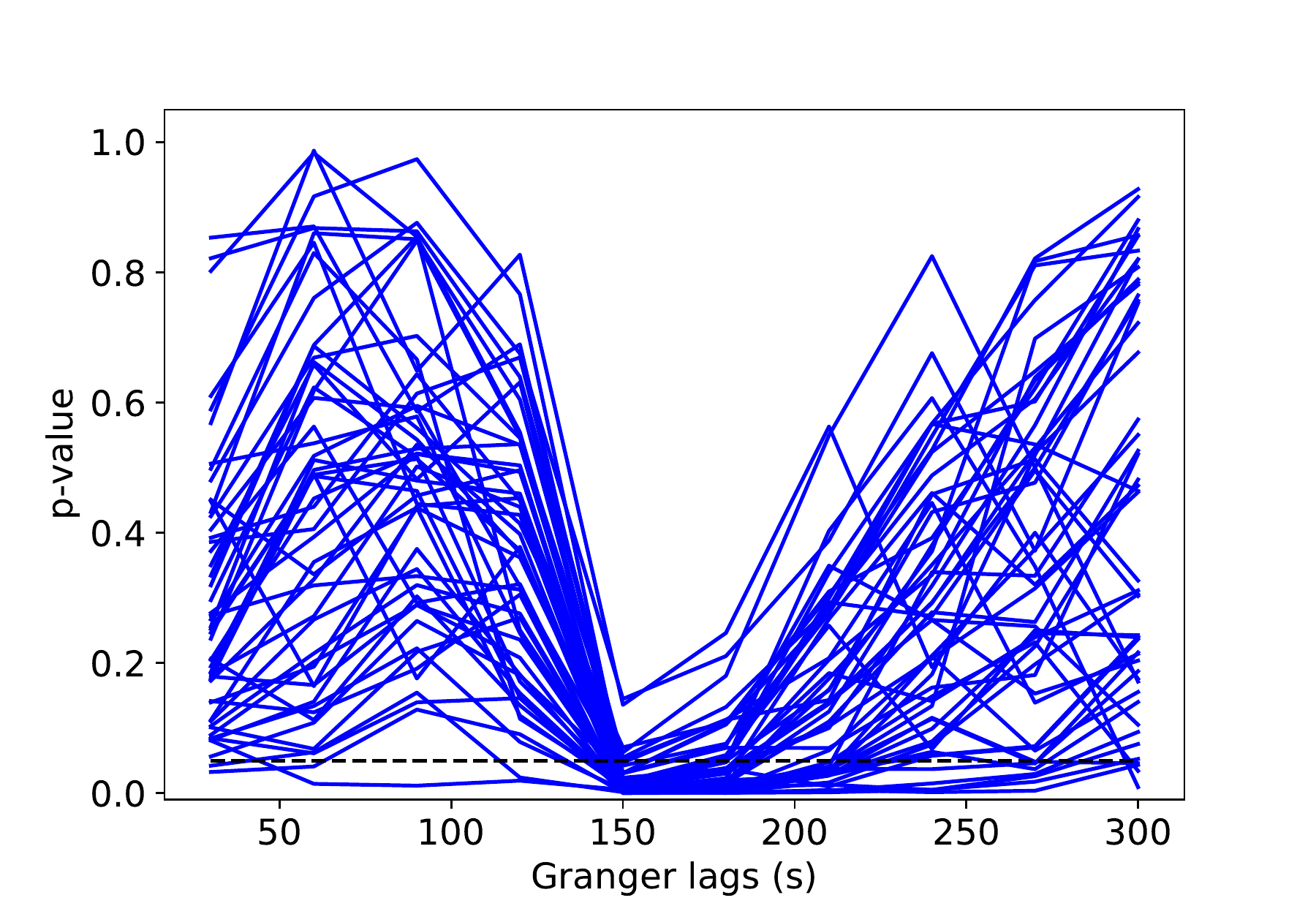}
\put(-117,143){$\sigma=0.3$, $\Delta t=30$~s}
}
\caption{Top panels: Granger-lag profiles ($h \rightarrow s$) varying with the size of the time bin, $\Delta t$, when uncorrelated variability with zero mean and a standard deviation of $\sigma = 0.1$ is added to the light curves. We fix $R=1$ and use a $\delta$-response function with $t_{\rm shift}=150$~s. For clarity, we show only 50 profiles which are randomly selected from 1,000 simulated profiles. Bottom panels: Same as the top panels, but for the case of $\sigma = 0.3$.}
\label{fig-random}
\end{figure*}

To estimate the effect of random noise, we use {\tt numpy.random.normal} to generate random numbers with normal distribution and further add them as uncorrelated variability to two light curves. The original light curves are generated with the mean of 0.5, while the uncorrelated variability has the mean of 0 and the standard deviation ($\sigma$) of 0.1 and 0.3. We simulate 1,000 events to see the variation of the Granger lag measurements induced by random noise in the light curves. The results are shown in Fig.~\ref{fig-random}. It can be seen that the noise or uncorrelated variability affects more on the significance of the lag than on the lag value. For example, the expected Granger-lag profiles can be confirmed in $\gtrsim 90 \%$ of the simulated samples that include uncorrelated variability with $\sigma=0.1$. Meanwhile, the error of the Granger lags, if observed using the procedure outlined in this work, likely depends on the chosen step size, or the size of the time bin ($\Delta t$).

\section{Effects of the interplay between the hard and soft X-ray lags on the Granger lags}
\label{sec:a2}

\begin{figure}
    \centerline{
        \includegraphics[width=0.5\textwidth]{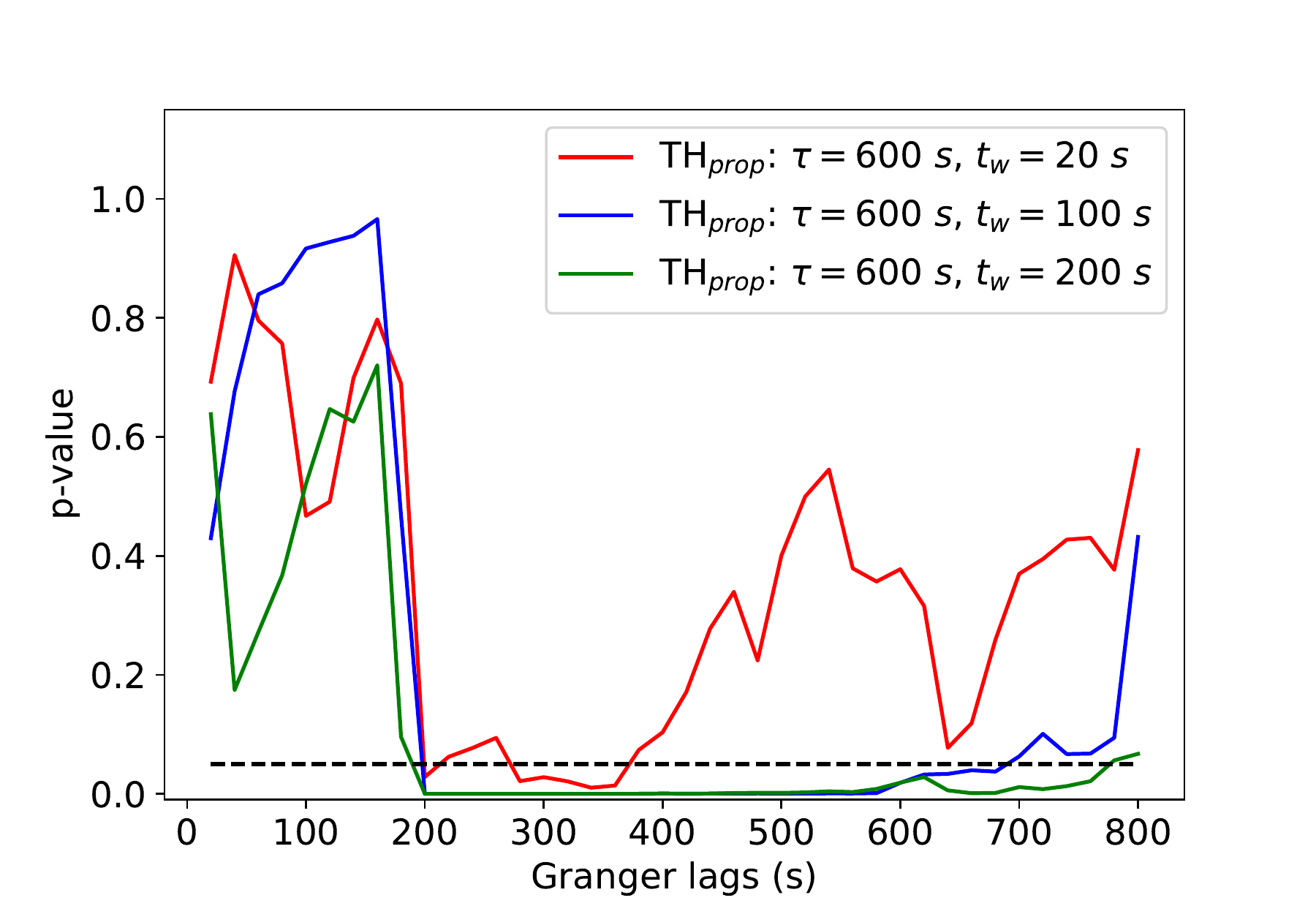}
    }
    \caption{Granger-lag profiles ($h \rightarrow s$) when two top-hat responses, TH$_\text{rev}$ and TH$_\text{prop}$, are included in the soft and hard bands, respectively, which are modelling reverberation and propagation lags. For TH$_\text{rev}$, we fix $\tau = 200$~s and $t_{w} = 20$~s. For TH$_\text{prop}$, we fix $\tau = 600$~s and vary $t_{w}$ to be 20, 100 and 200~s, in order to illustrate how the hard X-ray lags might affect the measured Granger lags.}

    \label{fig-B1}
\end{figure}

We investigate the presence of hard X-ray lags that might introduce a bias to the reverberation lags that are measured on shorter timescales, by including an additional top-hat response to the hard band. While the first top-hat added to the soft band represents the reverberation response to the direct X-ray continuum (TH$_\text{rev}$), the second top-hat function added to the hard band represents the response due to the disc propagating fluctuations (TH$_\text{prop}$). The result is shown in  Fig.~\ref{fig-B1}. We fix the centroid and the width of the TH$_\text{rev}$ to be $\tau = 200$~s and $t_{w} = 20$~s. On the other hand, the centroid of TH$_\text{prop}$ is fixed at $\tau = 600$~s while the width is varied to be $t_{w}= 20$, 100 and 200~s. All simulated light curves are binned with the bin size of $\Delta t = 20$~s. It can be seen that the hard lags do not significantly manifest the Granger lag measurements if the width of TH$_{prop}$ is larger than the width of TH$_{rev}$ (blue and green lines), which is likely the case for the propagation lags that dominate on longer timescales than the reverberation lags.

\bibliographystyle{mnras}

\bsp	
\label{lastpage}
\end{document}